\documentclass[10pt,journal,compsoc]{IEEEtran}
\ifCLASSOPTIONcompsoc
  \usepackage[nocompress]{cite}
\else
  \usepackage{cite}
\fi
             
\usepackage{soul}
\usepackage{float}
\usepackage{multirow}
\usepackage{epstopdf}
\usepackage[colorlinks,
            linkcolor=blue,       
            anchorcolor=blue,  
            citecolor=blue,        
            ]{hyperref}
\usepackage{listings}
\usepackage{fancybox}
\usepackage{graphicx}
\usepackage{subcaption}
\usepackage{paralist}
\usepackage{ragged2e}
\usepackage{enumitem}

\usepackage[framemethod=TikZ]{mdframed}
\usepackage{tcolorbox}
\makeatletter
\newcommand{\mybox}[1]{%
  \setbox0=\hbox{#1}%
  \setlength{\@tempdima}{\dimexpr\wd0+13pt}%
  \begin{tcolorbox}[boxrule=0.5pt, colback=white, arc=4pt,
      left=6pt,right=6pt,top=6pt,bottom=6pt,boxsep=0pt]
    #1
  \end{tcolorbox}
}
\usepackage{tikz}
\usepackage{pgfplots}
\usetikzlibrary{pgfplots.statistics,calc}
\usepackage{color}
\usepackage{xcolor}
\usepackage{array}
\usepackage{amsmath}
\usepackage{centernot}
\usepackage{xspace}
\usepackage{url}
\usepackage{verbatim}
\usepackage{wrapfig}
\usepackage{tabularx}
\usepackage{bbding}
\usepackage{pifont}
\usepackage{wasysym}
\usepackage{amssymb}

\newcommand{\tool}{\texttt{Nighthawk}}

\makeatletter  
\newif\if@restonecol  
\makeatother  
\usepackage[linesnumbered,ruled,vlined]{algorithm2e}
\usepackage{algpseudocode}  
\usepackage{amsmath}  

\begin{document}


\title{Nighthawk: Fully Automated Localizing \\ UI Display Issues via Visual Understanding}

\author{Zhe~Liu,
        Chunyang~Chen,
        Junjie~Wang,
        Yuekai~Huang,
        Jun~Hu,
        and~Qing~Wang
\IEEEcompsocitemizethanks{\IEEEcompsocthanksitem Zhe Liu is with the Laboratory for Internet Software Technologies, Institute of Software Chinese Academy of Sciences, University of Chinese Academy of Sciences, Beijing, China.
E-mail: liuzhe181@mails.ucas.ac.cn
\IEEEcompsocthanksitem Chunyang~Chen is with the Monash University, Melbourne, Australia.
E-mail: Chunyang.chen@monash.edu
\IEEEcompsocthanksitem Junjie Wang(corresponding author) and Qing Wang(corresponding author) are with the Laboratory for Internet Software Technologies, State Key Laboratory of Computer Sciences, Science \& Technology on Integrated Information System Laboratory, Institute of Software Chinese Academy of Sciences, University of Chinese Academy of Sciences, Beijing, China. 
E-mail: junjie@iscas.ac.cn, wq@iscas.ac.cn
\IEEEcompsocthanksitem Yuekai Huang and Jun Hu are with the Laboratory for Internet Software Technologies, Institute of Software Chinese Academy of Sciences, University of Chinese Academy of Sciences, Beijing, China.}

\thanks{Manuscript received April xx, xxxx; revised August xx, xxxx.}}

\markboth{IEEE TRANSACTIONS ON SOFTWARE ENGINEERING}%
{Shell \MakeLowercase{\textit{et al.}}: Bare Advanced Demo of IEEEtran.cls for IEEE Computer Society Journals}

\IEEEtitleabstractindextext{%
\begin{abstract}
Graphical User Interface (GUI) provides a visual bridge between a software application and end users, through which they can interact with each other.
With the upgrading of mobile devices and the development of aesthetics, the visual effects of the GUI are more and more attracting, and users pay more attention to the accessibility and usability of applications.
However, such GUI complexity posts a great challenge to the GUI implementation.
According to our pilot study of crowdtesting bug reports, display issues such as text overlap, component occlusion, missing image always occur during GUI rendering on different devices due to the software or hardware compatibility.
They negatively influence the app usability, resulting in poor user experience.
To detect these issues, we propose a fully automated approach, {\tool}, based on deep learning for modelling visual information of the GUI screenshot.
{\tool} can detect GUIs with display issues and also locate the detailed region of the issue in the given GUI for guiding developers to fix the bug.
At the same time, training the model needs a large amount of labeled buggy screenshots, which requires considerable manual effort to prepare them. 
We therefore propose a heuristic-based training data auto-generation method to automatically generate the labeled training data.
The evaluation demonstrates that our {\tool} can achieve average 0.84 precision and 0.84 recall in detecting UI display issues, average 0.59 AP and 0.60 AR in localizing these issues.
We also evaluate {\tool} with popular Android apps on Google Play and F-Droid, and successfully uncover 151 previously-undetected UI display issues with 75 of them being confirmed or fixed so far.
\end{abstract}

}

\maketitle

\IEEEdisplaynontitleabstractindextext

\IEEEpeerreviewmaketitle

\ifCLASSOPTIONcompsoc
\IEEEraisesectionheading{\section{Introduction}\label{sec_introduction}}
\else
\section{Introduction}
\label{sec_introduction}
\fi
\IEEEPARstart{G}{raphic} User Interface (GUI, also short for UI) plays an important role ubiquitous in almost all modern desktop software and mobile applications.
It provides a visual bridge between a software application and end users through which they can interact with each other.
Developers design a UI that requires proper user interaction, information architecture and visual effects of the UI.
Therefore, a good GUI design makes an application easy, practical and efficient to use, which significantly affects the success of the application and the loyalty of its users~\cite{jansen1998graphical}.

\begin{figure}[htb]
\centering
\vspace{-0.05in}
\includegraphics[width=8.8cm]{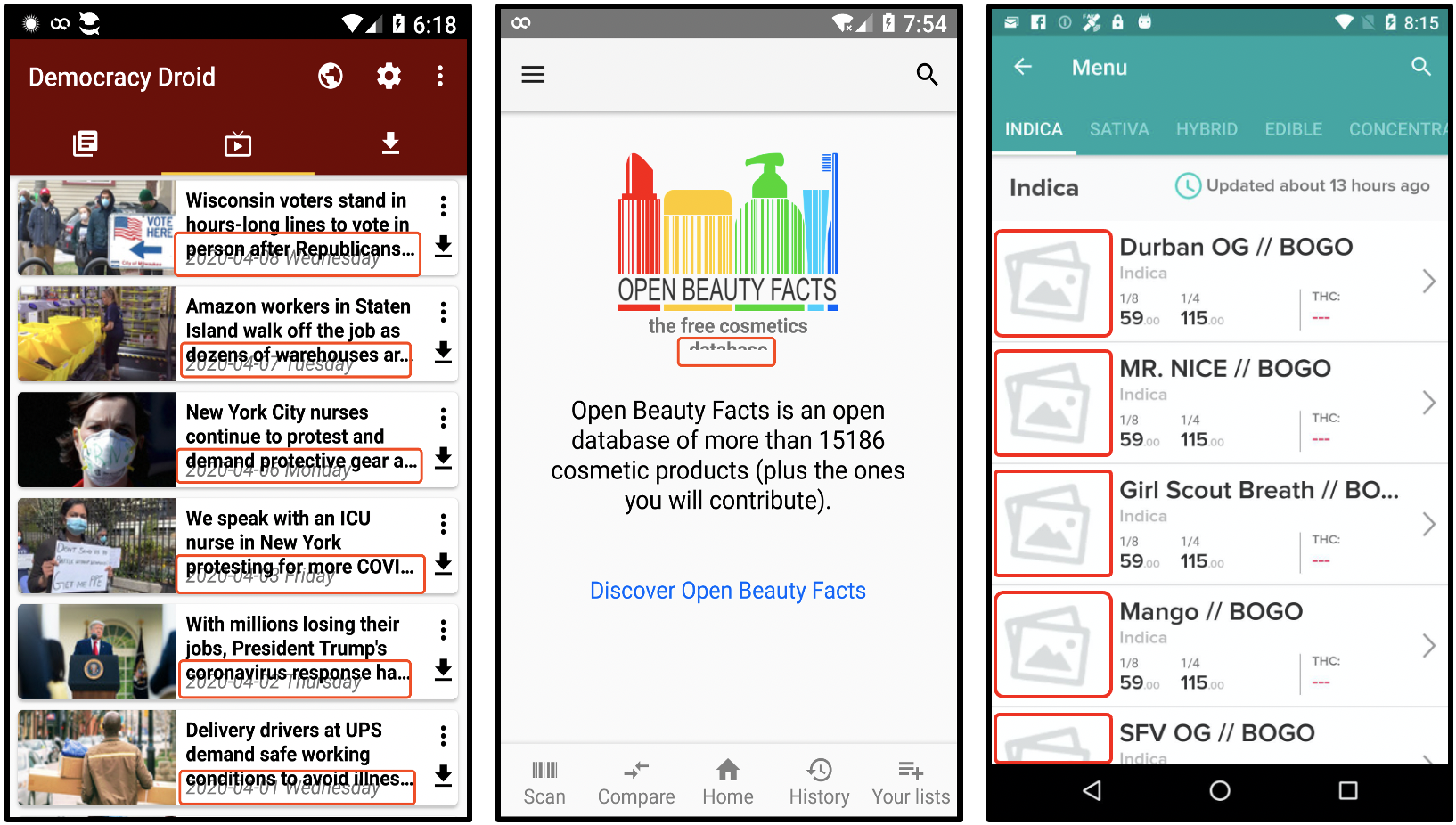}
\caption{Examples of UI display issues}
\label{fig:display issues}
\vspace{-0.05in}
\end{figure}

However, with the improvement of mobile device performance and user's aesthetic requirements for UI, more and more fancy visual effects in GUI design such as intensive media embedding, animation, light, floating and shadows post a great challenge for developers in the implementation.
Consequently, many display issues\footnote{we call these bugs as UI display issues, and will interchangablely use \textit{bug} and \textit{issue} in this paper.} 
such as \textit{text overlap, missing image, component occlusion} as seen in Figure~\ref{fig:display issues} always occur during the UI display process especially on different mobile devices~\cite{liu2022Guided,DBLP:conf/sigsoft/SuLC0W21}. 

In particular, we find that most of those UI display issues are caused by different system settings in different devices, especially for Android, as there are more than 10 major versions of Android OS running on 24,000+ distinct device models with different screen resolutions~\cite{wei2016taming}.
Although the software can still run along with these bugs, they negatively influence the fluent usage with the app, reduce the accessibility and usability of the application, resulting in the significantly bad user experience and corresponding loss of users.
Therefore, this study is targeting at detecting those UI display issues.

At present, to ensure the correctness of UI display, companies have to recruit many testers for app GUI testing or leverage the crowdtesting. 
Although human testers can spot these UI display issues, there are still two problems with such mechanism. 
First, it requires significant human effort as testers have to manually explore tens of pages by different interactive ways and also need to check the UI display on different OS versions and devices with different resolution or screen size. 
Second, some errors in the GUI display, especially relatively minor ones such as \textit{text overlap, component occlusion}, are often ignored by the testers.
To overcome those issues, some app development teams adopt the Rapid Application Development (RAD)~\cite{RAD}, which focuses on developing applications rapidly through frequent iterations and continuous feedback.
They utilize users' feedback to reveal UI display issues, but it is a reactive way for bug fixing which may have already hurt users of the app, resulting in the loss of market shares.

Therefore, in comparison with obtaining feedback from users for reactive app UI assurance, we need a more proactive mechanism which could check the UI display before the app release, automatically spot the potential issues in the GUI, and remind the developers to fix issues if any. 
There are many research works on automated GUI testing~\cite{mirzaei2016reducing, baek2016automated, su2017guided, 2017AutomatedGUI, Memon2013GUI, Coppola2017GUI, CrashScope2017Moran, Augusto2018M, GUI2019Denaro, Practical2019Gu, Paladin}
by dynamically exploring different pages with random actions (e.g., clicking, scrolling, filling in the text) until triggering the crash bugs or explicit exceptions.
Some practical automated testing tools like Monkey~\cite{Monkey, Wetzlmaier2017Hybrid}, Dynodroid~\cite{Dynodroid} are also widely used in industry.
However, these automated tools can only spot critical crash bugs, rather than UI display issues which cannot be captured by the system. 
In this work, we propose a method to automatically detect GUI display issues via visual understanding.

To have a preliminary understanding of the common UI rendering issues, we first carry out a pilot study on 10,330 non-duplicate screenshots from 562 mobile application crowdtesting tasks to observe display issues in these screenshots. 
Results show that a non-negligible portion (43.2\%) of screenshots are of display issues which can seriously impact the user experience, and degrade the reputation of the applications.
Besides, we also examine 1,432 screenshots from 200 random-chosen applications in the commonly-used Rico dataset ~\cite{Rico}, and find 8.8\% apps having UI display issues.
The common categories of UI display issues include \textit{component occlusion, text overlap, missing image, null value} and \textit{blurred screen}.
We use static analysis of XML file to analyze 100 screenshots (50 buggy screenshots and 50 bug-free screenshots) including component occlusion, text overlap and null value issues, since the missing image and blurred screen issues cannot be detected by static analysis of XML file. The results show that the precision of static analysis method is 0.32, and the recall is 0.36. 
The result shows that UI display issues cannot be completely detected only by static analysis.
Considering its popularity and lack of support in current practice of automatic UI testing, it would be valuable for classifying the screenshots with UI display issues from the plenty of screenshots generated during UI testing.

In our previous work, we proposed OwlEye~\cite{liu2020owl}, a UI display issue detection and localization approach. It adopts convolutional neural networks (CNN) to detect issues in screenshots, and uses Grad-CAM visualizes the issue location. 
However, OwlEye has two major limitations which seriously effected the practicality of widely deploying this method. First, the localized issue area by OwlEye is usually too large, thus can hardly accurately guide the follow-up bug fixing, especially the automatic bug fixing. 
Second, the OwlEye relies on large number of GUIs with display issues, and collecting the data is time-consuming.

To overcome the drawback of OwlEye and further improve its performance, this paper propose an extension, name {\tool}\footnote{Our approach is named {\tool} as it is like the nighthawk to effectively spot and localize UI display issues. And our model (catching small bugs at night like a nighthawk) can complement with conventional automated GUI testing (diurnal like eagle) for ensuring the robustness of the UI.}, to model the visual information by deep learning to automatically detect and localize UI display issues.
This paper formulate display issue detection as an object detection task. We adopt the Faster-RCNN model to not only identify screenshots with UI display issues but accurately point out the location of the visual issues within the screenshot, which better helps developers and testers debug their GUI code. 

Training the model needs large amount of buggy screenshots, which requires considerable manual effort to prepare them.
We therefore propose a heuristic-based training data auto-generation method to automatically generate the buggy screenshots. 
This is done through localizing and modifying the UI components related information (e.g., size of the TextView) in the JSON file of the bug-free screenshots.
Compared with our previous work, this method takes into account more restrictions (the text size and more components types) of the screenshots, so that the generated screenshots with issues are more realistic and diverse. Through user experiments, our latest issue data generated by {\tool} is closer to the real issue data than the data expansion method in OwlEye, and our data generated by {\tool} improves the performance of the model better than the method in OwlEye.

We train the model on 64,000 generated screenshots from 30,000 Android apps, and evaluate its effectiveness on 1,600 screenshots from crowdtesting and 8,000 augmented screenshots.
We first evaluated the performance of our method in bug detection. 
Compared with OwlEye and 13 other state-of-the-art baselines, our {\tool} can achieve more than 5\% and 6\% boost in recall and precision compared with OwlEye, and at least 17\% and 23\% boost in recall and precision compared with other baselines, resulting in 0.84 precision and 0.84 recall.
We further compare its localization results with OwlEye. The average precision and average recall of {\tool} are 55\% and 56\% higher than those of OwlEye, with AP of 0.59 and AR of 0.60.

Apart from the accuracy of our {\tool}, we also evaluate the usefulness of our {\tool} by applying it in detecting the UI display issues in the real-world apps from Google play and F-Droid.
Among 1328 apps, we find that 151 of them are with UI display issues.
We reported bug reports to the development team and 75 are confirmed and fixed by developers.

\begin{figure*}[htb]
\centering
\includegraphics[width=18.0cm]{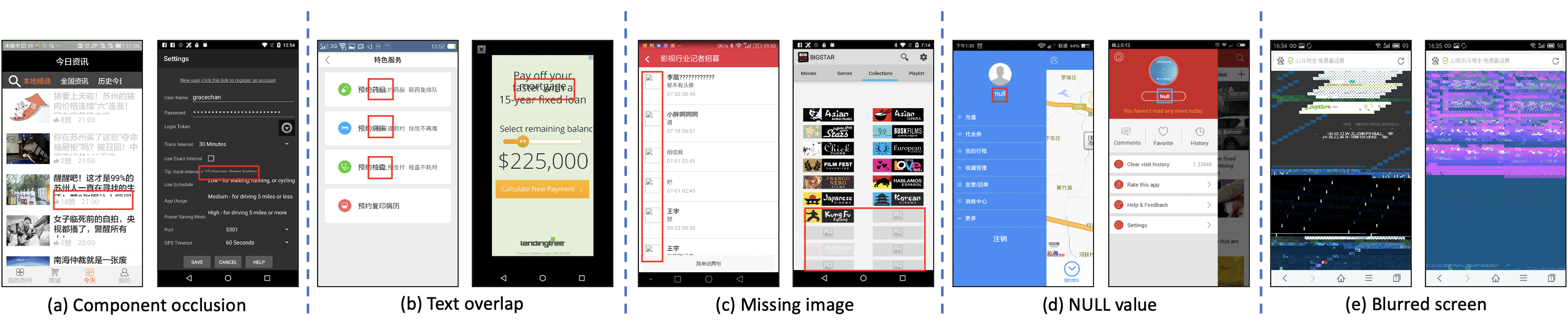}
\caption{Examples of five categories of UI display issues}
\label{fig:5-kind-bug}
\vspace{-0.15in}
\end{figure*}

This paper is an extended version of our earlier study~\cite{liu2020owl}. The extension makes the following additional contributions:

\begin{itemize}

\item We adopt the Faster-RCNN model to not only identify buggy GUI screenshot, but accurately point out the location of display issues within the screenshot, which better helps developers and tester debug their GUI code.

\item To avoid the requirement of large-scale manual labeling, we propose a heuristic-based training data auto-generation method to automatically generate diverse and realistic screenshots with UI display issues from bug-free UI screenshots.

\item We carry out experiments on a large-scale dataset to verify that the {\tool} can automatically train and detect UI display issues without manual annotation with promising results.
At the same time, we also evaluate the issue localization performance of the {\tool} and the impact of the number of datasets on the performance of the {\tool}.

\item We release the implementation of {\tool}\footnote{https://github.com/20200501/Nighthawk.\label{github}}, the detailed experimental results, and the large-scale dataset of app UIs with four kinds of issues and issue localization information, for other researchers' follow-up studies.

\end{itemize}

\section{motivational study}
\label{sec_motivation}
In order to get a better understanding of the UI displaying issues in real-world practice, we carry out a pilot study to examine the prevalence of these issues. 
The pilot study also explores what kinds of UI display issues exist, so as to facilitate the design of our approach for detecting UIs with display issues.

\subsection{Data Collection}
\label{sec_motivation_data_collection}

Our experimental dataset is collected from one of the largest crowd-testing platforms\footnote{Baidu (baidu.com) is the largest Chinese search service provider.
Its crowdsourcing test platform (test.baidu.com) is also the
largest ones in China.} in which crowd workers are required to submit test reports after performing testing tasks~\cite{wang2019iSENSE,wang2020context,wang2021context}.
The dataset contains 562 Android mobile application crowdtesting tasks between January 2015 and September 2016.
These apps belong to different categories such as news, entertainment, medical, etc. 
In each task, crowd workers submit hundreds of testing reports which describe how the testing is conducted and what happened during the test, as well as accompanied screenshots of the testing. 
The reason why we utilize this dataset is that it includes both the UI screenshots and the corresponding bug description which facilitates the searching and analysis of UI display issues.
This dataset contains 10,330 unique GUI screenshots.

\subsection{Categorizing UI Display Issues}
\label{sec_motivation_types_UIs}

For these GUI screenshots, the first three authors individually check each of them manually with also its corresponding description in the bug report.
Only GUI screenshots with the consensus from all three human markers are regarded as ones with display issues. 
A total of 4,470 GUI screenshots are determined with UI display issues, which accounts for 43.2\% (4470/10330) in all screenshots.
This result indicates that the UI display issues account for a non-negligible portion of mobile application bugs revealed during crowdtesting and should be paid careful attention for improving the software quality.

During the manual examination process, we notice that there are different types of UI issues, a categorization of these issues would facilitate the design and evaluation of related approach.
The first and the third authors manually check it following the open coding procedures~\cite{DBLP:journals/tse/Seaman99}. We analyze the issue screenshots and categorize the UI display issues.
In detail, each annotator carefully examines the issue screenshots. 
We group similar codes into one category, and the grouping process is iterative. 
Specifically, we constantly move back and forth between the category. In the absence of an agreement between the two authors, the second author act as arbitrators to discuss and resolve the conflict.
We follow the procedure until all authors reach an agreement. 
We classify those UI issues into five categories including \textit{component occlusion, text overlap, missing image, null value} and \textit{blurred screen} with details as follows:

\textbf{Component occlusion (47\%)}: 
As shown in Figure \ref{fig:5-kind-bug}(a), 
the textual information or component is occluded by other components. 
It usually appears together with TextView or EditText.
The main reasons are as follows: the improper setting of element's height, or the adaptive issues triggered when setting a larger-sized font.

\textbf{Text overlap (21\%)}: 
As shown in Figure \ref{fig:5-kind-bug}(b), two pieces of text are overlapped with each other. 
This might be caused by the adaptive issues among different device models, e.g., when using a larger-sized font in a device model with small screen might trigger this bug.

Note that, for text overlap category, two pieces of text are mixed together; while for component occlusion, one component covers part of the other component.

\textbf{Missing image (25\%)}: 
As shown in Figure \ref{fig:5-kind-bug}(c), in the icon position, the image is not showing as its design.
The possible reasons are as follows: wrong image path or layout position, unsuccessful loading of the configuration file due to permissions, oversized image, network connection, code logic, or picture errors, etc.

\textbf{NULL value (6\%)}: 
As shown in Figure \ref{fig:5-kind-bug}(d), the right information is not displaying, instead \textit{NULL} is showing in corresponding area.
This category of bugs usually occurs with TextView. 
The main reasons are as follows: issues in parameter setting or database reading, and the length of text in TextView exceeding the threshold, etc.

\textbf{Blurred screen (1\%)}: As shown in Figure \ref{fig:5-kind-bug}(e), the screen is blurred.
The reason for this bug might because the defects in hardware, or the exclusion of hardware acceleration for some CPU- or GPU- demanding functionalities. 

To further validate the generality of our observations, we also manually check 1,432 screenshots from 200 random-chosen applications in Rico\footnote{http://interactionmining.org/rico\#quick-downloads} dataset~\cite{Rico} , which is a commonly-used mobile application dataset with 66K UI screenshots of Android Applications and we will further introduce that dataset on Section~\ref{subsec_approach_augmentation}.
We find that 18 UIs from 16 apps (16/200 = 8.8\% apps) are with UI display issues.
Note that number is highly underestimated, as the collected UIs do not cover all pages of the applications, and the applications are not fully tested on different devices with different screen resolutions.

\subsection{Why Visual Understanding in Detecting UI Display Issues}
\label{sec_motivation_limitations}

These findings confirm the severity of UI display issues, and motivate us to design approach for automatically detecting these GUI issues.
One commonly-used practice for bug detection in mobile apps is the program analysis, but it may not be suitable in this scene. 
To apply the program analysis, one need to instrument the target app, develop different rules for different types of UI display issues, rewrite the code for different platforms (e.g., iOS, Android), and customize their code to be compatible on different mobile devices (e.g., Samsung, Huawei, etc) with different screen resolution, which is extremely effort-consuming. 
Specifically, it is not trivial to enumerate all display issues and develop corresponding rules for detection.

\begin{figure}[htb]
\centering
\vspace{-0.05in}
\includegraphics[width=8.6cm]{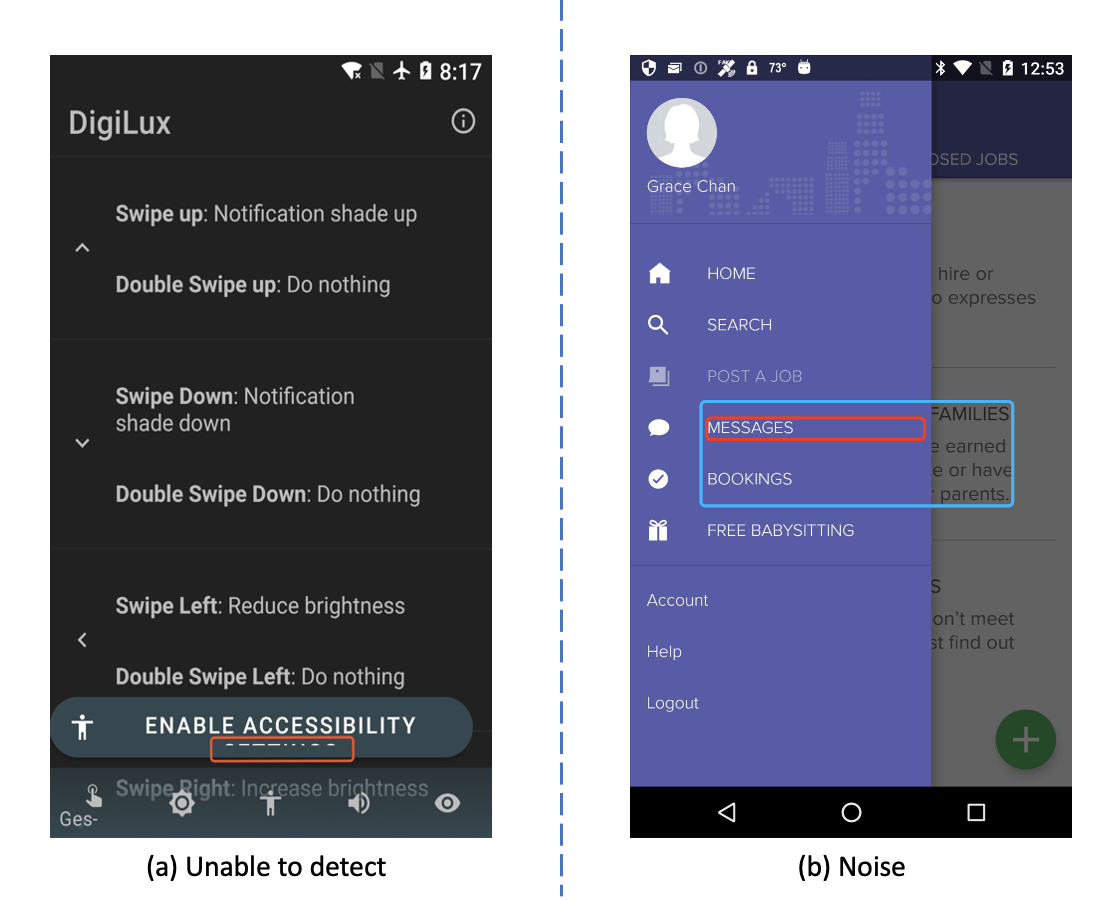}
\caption{Examples of unable to detect and noise}
\label{fig:noise}
\vspace{-0.05in}
\end{figure}

We observed the screenshots of these 5 categories of bugs and their corresponding JSON files, and found that component occlusion, text overlap and null value can be detected by analyzing JSON files to obtain component information(i.e., component coordinates, component text). 
We try to use the static analysis of XML files to detect these three categories of bugs as follows.
For component occlusion, we analyze the coordinates of all components, and regard the components with intersection coordinates as bugs. 
For text overlap, we analyze the coordinates of all text views, and regard the components with intersection coordinates as bugs. 
For null value, we analyze the text of the component, and regard null text as a bug.

However, static analysis method has some limitations. In the following cases, it is impossible to detect bugs by analyzing JSON files. 
For component occlusion, as shown in Figure~\ref{fig:noise}(a), because the font size in the JSON file cannot be obtained, the issue that the font is displayed incompletely in EditText cannot be detected. 
As shown in Figure~\ref{fig:noise}(b), the toolbar, spinner and dialog will float in front of the component, which will cause noise to the detection. 
For text overlap, there is still noise such as component occlusion. 
For null value, there are more nulls in the text acquisition process, most of which are due to the problems existing in the process of getting JSON files, but there is no problem in the actual UI display, which will add a lot of noise to the detection of such bugs.

Taken in this sense, it is worthwhile developing a new efficient and general method for detecting UI display issues.
Inspired by the fact that these display issues can be spotted by human eyes, we propose to identify these buggy screenshots with visual understanding technique which imitates the human visual system.
As the UI screenshots are easy to fetch (either manually or automatically) and exert no significant difference across the apps from different platforms or devices, our image-based approach are more flexible and easy to deploy.

\begin{figure*}[htb]
\centering
\vspace{0.3in}
\includegraphics[width=18.0cm]{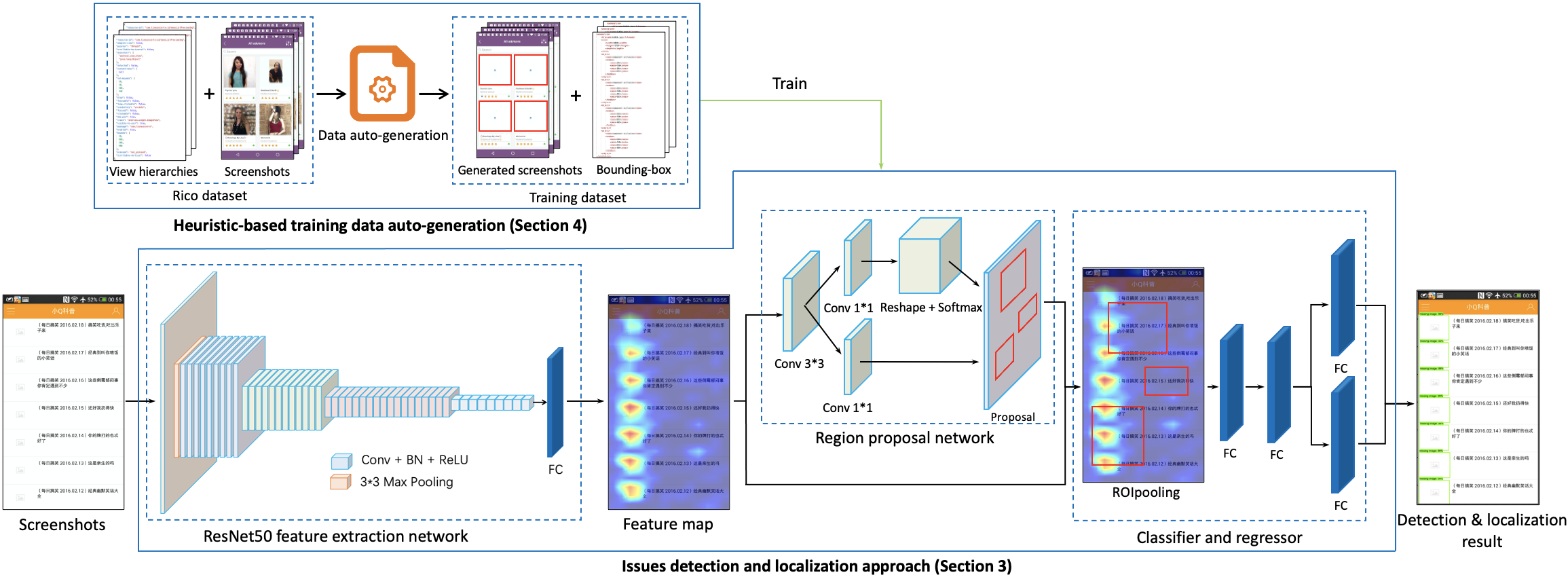}
\caption{Overview of {\tool}}
\label{fig:overview}
\vspace{-0.1in}
\end{figure*}

\section{Issues Detection and Localization Approach}
\label{sec_approach}

This paper proposes {\tool} to automatically detect and localize UI display issues in the screenshots of the application under test, as shown in Figure \ref{fig:overview}.
Given one UI screenshot, our {\tool} provides integrated detection and localization services. {\tool} can detect the screenshot related with UI display issues via visual understanding and localize the detailed issue position by bounding boxes on the UI screenshot for guiding developers to fix the bug.

As the UI display issues can be spotted via the visual information, we adopt the Faster-RCNN~\cite{ren2015faster}, which has proven to be effective in object detection in computer vision domain. Figure \ref{fig:overview} shows the structure of our object detection model which includes a feature extraction network (we use ResNet50), a regional proposal network (RPN) module, and an ROI pooling module. 

Given the input UI screenshot, we convert it into a certain image size with fixed width and height as $w \times h$, and the image is normalized. Then the screenshot is input into the convolution neural network ResNet50~\cite{he2016deep}. The Convolutional layer's parameters consist of a set of learnable filters. The purpose of the convolutional operation is to extract the different characteristics of the input (i.e., feature extraction). After the convolutional layer, the screenshots will be abstracted as the feature graph. 
However, with the network depth increasing, accuracy gets saturated and then degrades rapidly, it is easy to appear the vanishing / exploding gradients problem and degradation problem. Because the gradient propagates back to the previous layer, repeated multiplication may make the gradient infinitesimal. As a result, with the deeper layers of the network, its performance tends to be saturated or even rapidly decline. In order to solve this problem, ResNet50 introduces the concept of residual error to solve this problem. ResNet50 solves the degradation problem by introducing a deep residual learning framework. Instead of making each layer directly fit a desired underlying mapping, it explicitly matches these layers with a residual mapping. 

After obtaining the feature map through ResNet50, we input the feature map into Region Proposal Network (RPN) module. The RPN takes the feature map (of any size) as input and outputs a set of rectangular 3 object proposals, each with an objectness score. Then, a 3x3 slide window is used to traverse the whole feature map. In the process of traversing, nine anchors are generated according to rate and scale (1:2, 1:1, 2:1) in each window center. Then, full connection is used to classify each anchor (foreground or background) and preliminary bounding boxes expression. Then the bounding box expression is used to modify the anchors to obtain accurate proposals. 

According to the feature map obtained by the feature extraction module and proposal obtained by RPN module. Input it into the ROI pooling layer to calculate the proposal feature maps. Finally, the proposal feature maps are input into the classification module, and the specific category (such as component occlusion, missing image, etc.) of each proposal is calculated through the fully connected neural networks (FC) and softmax layer, and the probability vector is output. At the same time, the position offset of each proposal is obtained by using bounding box expression again, which is used to regression more accurate target detection frame.

\section{heuristic-based training data auto-generation}
\label{subsec_approach_augmentation}

Training an object detection model for visual understanding requires a large amount of input data.
For example, ResNet~\cite{ResNet} model uses 128 million images from ImageNet as training dataset for image classification task.
Similarly, training our proposed Faster-RCNN for UI display issues detection and localization requires abundant of screenshots with UI display issues.
However, there is so far no such type of open dataset, and collecting the related buggy screenshots is quite time- and effort-consuming. 
Different from image classification task, Faster-RCNN model not only needs to determine whether there are bugs on the screenshots, but also to mark the specific location of bugs on the screenshots. This requires a large number of experienced testers to mark it.
At the same time, the approach in our previous work OwlEye needs a certain number of screenshots with UI display issues, and collecting these issue screenshots requires human annotation, since most screenshots are issue-free.
Therefore, we develop a heuristic-based training data auto-generation method for generating UI screenshots with UI display issues from bug-free UI images.

The data auto-generation is based on the Rico~\cite{Rico}
dataset which contains more than 66K unique screenshots from 9.3K Android applications, as well as their accompanied JSON file (i.e., detailed run-time view hierarchy of the screenshot). 
According to our observation on Section~\ref{sec_motivation}, most UI screenshots in this dataset are of no dispaly issues.

\begin{figure}[htb]
\centering
\vspace{0.2in}
\includegraphics[width=8.8cm]{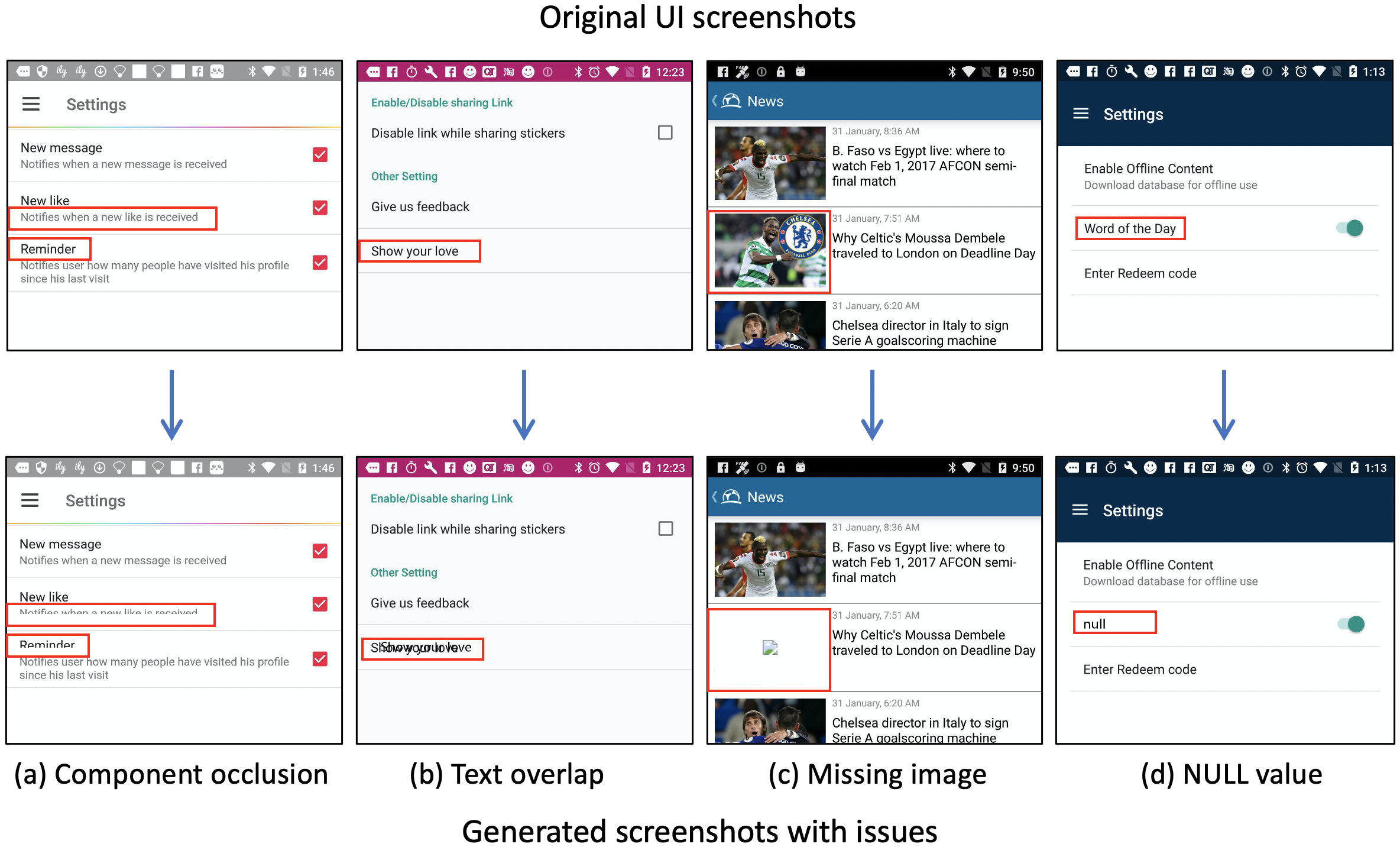}
\caption{Examples of data auto-generation}
\label{fig:data_aug-4}
\vspace{-0.1in}
\end{figure}

Algorithm 1 presents the heuristic-based training data auto-generation algorithm.
With the input screenshot and its associated JSON file, the algorithm first locates all the TextView and ImageView, then randomly chooses a TextView or ImageView depending on the generated category.
Based on the coordinates and size of the TextView/ImageView, the algorithm then makes its copy and adjusts its location or size following specific rules to generate the screenshot with corresponding UI display issues (line 1-12).  
Figure \ref{fig:data_aug-4} demonstrates the illustrative examples of the generated screenshots with UI display issues.

Note that, among the five categories of UI display issues, the category of \textit{blurred screen} is difficult to generate following the above idea. 
In addition, the preliminary survey results show that the number of such bugs is small, accounting for only 1\% of the crowdtesting data (see Section \ref{sec_motivation_types_UIs}).
Hence, we leave this category for future work.
For the generation methods of these four kinds of issues, we conducted a pilot study. 
In detail, we select 500 issue screenshots of different apps for each type of issue to summarize the characteristics of issue screenshots, and the settings and parameters of the auto-generation method is based on the summarized characteristics.
The general principle of the auto-generation is that we randomly decide the occlusion/overlap offset or occlusion/overlap region to generate diversified screenshots, and we assume the large number of generated screenshots in training dataset can mitigate the problem caused by too large or too slight offset.
We then present the detailed auto-generation rules of the four categories.
\begin{algorithm}  
  \caption{Heuristic-based training data auto-generation}  
  \KwIn{
       $scr$: screenshot without bugs\; 
       $json$: associated JSON file\;
       $category$: category of generated UI display issue\;
       $icon$: pre-prepared image icon\;
       }  
  \KwOut{
        $genscr$: generated screenshot with \textit{category} bug\;
        $bbox$: coordinates of bounding box\;
        }  
  
  Traverse $json$ file to obtain all TextView \& ImageView \& Button \& EditText\;
  
  \If{$category==$`$missing \ image$'} 
  {
    Randomly choose an ImageView\; 
  }
  \If{$category==$`$component \ occlusion$'} 
  {
    Randomly choose a TextView / ImageView / Button / EditText\; 
  }
  \If{$category==$`$text \ overlap$' $or$ `$null \ value $'} 
  {
    Randomly choose a TextView\;
  }
  
  Obtain the coordinates of TextView / ImageView / Button / EditText  ($x_1$,$y_1$),($x_2$,$y_2$); \
  \textcolor{gray}{//coordinate of upper left and lower right (If it is textview, get the upper and lower left coordinates of text)}\ 

  Calculate the width and height of TextView / ImageView / Button / EditText($w$, $h$) based on the coordinates\;
 
  Obtain the text content of TextView ($text$)\;
  
  Obtain the text font size of TextView ($fs$)\;
   
  Obtain the background color of TextView / ImageView / Button / EditText ($bg$)\;
  
  \If{$category==$`$component \ occlusion$'}
  {
    $rand\gets$ random.uniform(-1,1)\;
    $image$.new($(w, h\times \vert rand\vert))$,$bg$,$fs$)\;
    \If{$rand \geq 0$}
    {
    \textcolor{gray}{//Occlude the upper part of component}
        $genscr\gets scr$.paste($image$, ($x_1$,$y_1$))\;
    }
    \Else{
    \textcolor{gray}{//Occlude the lower part of component}
    $genscr\gets scr$.paste($image$, ($x_1$,$y_2+(h\times rand)$));}
  }
  \If{$category==$`$text \ overlap$'}
  {
    $xrand\gets$ random.uniform($-0.5\times w,0.5\times w$)\;
    $genscr\gets scr$.write([$x_2-xrand,y_1$],$text$,$fs$)\;
    \textcolor{gray}{//Get the coordinates of the overlap}
    $x_1$,$y_1$,$x_2$,$y_2\gets$getoverlap($genscr$,$fs$)\;
  }
  \If{$category==$`$missing \ image$'}
  {
    $image$.new(($w,h$),$bg$)\;
    $scr$.paste($image$,($x_1$,$y_1$))\;
    $genscr\gets scr$.paste($icon$, ($x_1+0.5\times w$, $y_1+0.5\times h$))\;
  }
  \If{$category==$`$null \ value$'}
  {
    $image$.new(($w,h$),$bg$)\;
    $scr$.paste($image$, ($x_1$,$y_1$))\;
    $genscr\gets scr$.write([$x_1$,$y_1$],``null'',$fs$)\;
  }
  $bbox\gets$ writexml($x_1$,$y_1$,$x_2$,$y_2$)\;
  return $genscr,bbox$\;  
\end{algorithm}  


\textbf{Auto-generation for Component Occlusion Bug}: 
When this category of bug occurs, the textual information or component is occluded by other components.
In detail, the pilot study reveals that it appears in TextView, EditText, Button or ImageView, and the auto-generation is conducted on them randomly.
The pilot study also reveals that when occlusion happens, the two involved components are usually with the same width but different height.
Therefore, we first generate a color block with the same width and background color as the component but with a smaller height (i.e., randomly-generated value), and then put it to cover part of the component randomly (e.g., lower left part).
The random generated height is determined by $random(-1,1)$ guided by the pilot study.
For the determination of the color of the occlusion area, we obtain the color of the upper left corner and the upper right corner of the component, and take the average as the color of the occlusion area.
Finally, we get the size of the component as the bounding box (line 13-19).

\textbf{Auto-generation for Text Overlap Bug}: 
The textual contents are overlapped with each other, when this category of bug occurs.
To auto-generate this category of screenshots, we first get the height of the TextView, convert it to a font size, and generate a piece of text with the same size and content as the original TextView, and offset it slightly. The threshold of offset is calculated by random number, that is, $random(-0.5\times w,0.5\times w)$ which is observed in the pilot study. Besides, through the pilot study, we determined that the color of the font is also randomly selected (black, gray, white).
Finally, we use the overlapped part of the offset text and the original text as the bounding box (line 20-23).

\textbf{Auto-generation for Missing Image Bug}: 
We notice that when this category of bug occurs, an image icon would show up to indicate that the area supposes to be an image.
We summarize 10 common icons from our pilot study. These icons are quite different from real-world images, and these issue image icons are rarely used in the UI of app (most UI developers think these icons are issue icons), so the detection results are mostly issue. 
To auto-generate this category of screenshots, we first download 10 frequently-used image icons online, then cover the original image displaying area with one random-chosen image icon and set its background color as the color of its original image. 
Through our real-world app detection in RQ4, we find there are slight different icons which can be detected by {\tool} (indicating the effectiveness of our propose approach), and donot find entirely different new icons (indicating the relatively completeness of the training dataset.). Finally, we use the region of ImageView as the bounding box. 
 (line 24-27).

\textbf{Auto-generation for NULL Value Bug}: 
When this category of bug occurs, \textit{NULL} is displayed in the area where supposes to be a piece of text. 
We first get the height of the text in TextView and convert it to font size. Then we generate this category of screenshots by covering the original TextView using a color block which shares the same background color and adds \textit{NULL} with the same font size as the original text. Finally, use the text area of \textit{NULL} as the bounding box (line 28-31).

Note that, both \textit{component occlusion} and \textit{text overlap} involves covering a TextView, the difference is that the former one utilizes a color block to cover the TextView so that it looks like a component blocks the text, while the latter one employs a piece of text to cover the TextView to make it look like the two pieces of text are overlapped with each other.
Another note is that, based on our observation on the screenshots with UI display issues in Section \ref{sec_motivation}, when conducting the auto-generation, the TextView is covered in the vertical direction in \textit{component occlusion}, while it is covered in the horizontal direction in \textit{text overlap}.

\section{Experiment Design}
\label{sec_exeperiment}

\subsection{Research Questions}
\label{subsec_experiment_RQ}
\begin{itemize}[leftmargin=*]
\item \textbf{RQ1: (Issues Detection Performance)} How effective is our proposed {\tool} in detecting UI display issues? 
\end{itemize}

For RQ1, we first present some general views of our proposed approach for UI display issues detection and the comparison with commonly-used baseline approaches (details are in Section \ref{subsec_experiment_baseline}).  

\begin{itemize}[leftmargin=*]
\item \textbf{RQ2: (Issues Localization Performance)} How effective is our proposed {\tool} in localizing UI display issues?
\end{itemize}

For evaluating the performance of issues localization, we compare it with our previous approach OwlEye, and detect its accuracy through its average recall (AR) and average precision (AP).

\begin{itemize}[leftmargin=*]
\item \textbf{RQ3: (Contribution of Data Auto-generation)} What is the contribution of the data auto-generation approach?
\end{itemize}

For RQ3, we evaluate the contribution of data auto-generation. 
We first examine the influence of auto-generated training data size on the model performance, then experimentally compare the issue detection performance between the model with this upgraded auto-generation approach and the model with the augmentation approach of OwlEye.

\begin{itemize}[leftmargin=*]
\item \textbf{RQ4: (Usefulness Evaluation)} How does our proposed {\tool} work in real-world situations?
\end{itemize}

For RQ4, we integrate {\tool} with DroidBot as a fully automatic tool to collect the screenshots and detect UI display issues, and then issue the detected bugs to the development team.

\subsection{Experimental Setup}
\label{subsec_experiment_dataset}

As our {\tool} is a fully automatic approach, we use the heuristic-based training data auto-generation approach in Section \ref{subsec_approach_augmentation} to generate a large number of data.
In detail, we randomly download one or two screenshots from each of the random-chosen 30,000 applications in Rico dataset, and each screenshot would be utilized once for the data auto-generation. 
In order to make the training data balanced across categories, we used the same number of screenshots as training data for each type of issues.

For the auto-generated 50,000 screenshots with UI display issues from Alg 1, we first extract their features with ORB feature extraction algorithm~\cite{ORB}, rank them randomly, compute the cosine similarity between a specific screenshot and each of its previous ones, and remove it when a similarity value above 0.8 is observed.
In this way, 40,000 screenshots with UI display issues from 50,000 screenshots (each category has 10,000 screenshots) and equal number of bug-free non-duplicate screenshots (from buggy screenshots corresponding bug free versions), with a total of 80,000 screenshots are remained and added into the experimental dataset.

\begin{table}[H]
\caption{The number of 5 categories of buggy screenshots}
\vspace{-0.05in}
\label{tab:5-kind-buggy-screenshots}
\centering
\footnotesize
\begin{tabular}{p{2.8cm}<{\centering}| p{0.65cm}<{\centering}| p{0.8cm}<{\centering} | p{1.2cm}<{\centering}|p{0.6cm}<{\centering}}
\hline
\multirow{2}{*}{\textbf{Category}} & \multirow{2}{*}{\textbf{Train}} & \multicolumn{2}{c|}{\textbf{Test}} & \multirow{2}{*}{\textbf{Val}} \cr\cline{3-4}

 & & \textbf{Crowd} & \textbf{AutoGen} \\
\hline
Component occlusion & 8000 & 200 & 1000 & 1000\\ 

Text overlap & 8000 & 200 & 1000 & 1000\\ 

Missing image & 8000 & 200 & 1000 & 1000\\

NULL value & 8000 & 200 & 1000 & 1000 \\ 

\hline
\hline
Overall & 32000 & 800 & 4000 & 4000\\

\hline

\end{tabular}
\vspace{-0.05in}
\end{table}

In order to simulate the real-world application of our proposed {\tool}, we setup the experiment as follows. 
For the 80,000 screenshots of the Rico dataset (the ratio of positive and negative samples is 1:1), the 40,000 screenshots with UI display issues are generated as positive samples for the experiment, including 10,000 screenshots for each of the four categories of bugs (Each category selects the same number of negative samples). 
Set the training set, testing set and validation set according to the ratio of 8:1:1. 
According to the same experimental setup, always ensure that the ratio of positive and negative samples is 1:1, 1,000 buggy screenshots were randomly selected as the testing set (Only used in the issues detection performance by category in Section \ref{sec_results_RQ1}), 1,000 buggy screenshots as the validation set, and the remaining 8,000 buggy screenshots as the training set.

In addition, in order to understand the performance of the {\tool} in real-world dataset, the testing set also includes the 1,600 screenshots (800 with UI display issues and 800 without) from 300 crowdtesting apps (Note that in order to compare with OwlEye, we use the same test dataset in OwlEye.), we utilize the 400 screenshots (200 with UI display issues and 200 without) 400 screenshots for each category of bugs as testing set (Used in Section \ref{sec_results_RQ1}-\ref{sec_results_RQ2}) to evaluate the performance of {\tool}.
Considering the long training time, we used the 3-fold cross-validation. For simplicity, we present the average performance of the experimental results.

Table \ref{tab:5-kind-buggy-screenshots} presents the distribution of screenshots in terms of different categories.
The model is trained in a NVIDIA GeForce RTX 2060 GPU (16G memory) with 100 epochs for about 8 hours.

\subsection{Baselines}
\label{subsec_experiment_baseline}
To further demonstrate the advantage of {\tool}, we compare it with 6 baselines utilizing both machine learning and deep learning techniques.
The 4 machine learning approaches first extract visual features from the screenshots, and employ machine learners for the classification.
The 2 deep learning approaches utilizes artificial neural network directly on the screenshots for classification.
We first present the feature extraction approach used in machine learning approaches.

\textbf{SIFT}~\cite{SIFT}: Scale invariant feature transform (SIFT) is a common feature extraction approach to detect and describe local features in an image. It can extract the interesting points on the object to generate the feature description of the object, which is invariant to uniform scaling, orientation, and illumination changes.

\textbf{SURF}~\cite{SURF}: Speed up robot features (SURF) is an improvement of \textit{SIFT}.
SURF uses an integer approximation of the determinant of Hessian blob detector, which can be computed with 3 integer operations using a precomputed integral image.

\textbf{ORB}~\cite{ORB}: Oriented fast and rotated brief (ORB) is a fast feature point extraction and description algorithm. Based on the rapid binary descriptor ORB of brief, it has rotation invariance and anti noise ability.

With these features, we apply four commonly-used machine learning approaches, i.e., \textbf{Support Vector Machine (SVM)}~\cite{kotsiantis2007supervised}, \textbf{K-Nearest Neighbor (KNN)}~\cite{berson2004overview}, \textbf{Naive Bayes (NB)}~\cite{kotsiantis2007supervised} and \textbf{Random Forests (RF)}~\cite{breiman2001random}, for classifying the screenshots with UI display issues. 

\textbf{MLP}~\cite{Deeplearning, Deeplearningnature}: Multilayer Perceptron (MLP) is a feedforward artificial neural network. The network structure is divided into input layer, hidden layer and output layer. Each node is a neuron that uses a nonlinear activation function, e.g., corrected linear unit (ReLU). It is trained by changing the connection weight according to the output error compared with the ground truth. We used eight layers of neural network, and we set the number of neurons in each layer to 190, 190, 128, 128, 64, 64, 32 and 2, respectively.

\textbf{OwlEye}~\cite{liu2020owl}: OwlEye builds on the Convolutional Neural Network (CNN) to identify the screenshots with UI display issues, and utilizes Gradient weighted Class Activation Mapping (Grad-CAM) to localize the regions with UI display issues.

\subsection{Evaluation Metrics}
\label{subsec_experiment_evaluation_metrics}

In order to evaluate the issues detection performance of our proposed approach in RQ1, we employ three evaluation metrics, i.e., precision, recall, F1-Score, which are commonly-used in image classification and pattern recognition ~\cite{Easy-to-Deploy,manning2008introduction}.
For all the metrics, higher value leads to better performance.

Precision and recall are often calculated by counting true positions (TP), true negatives (TN), false positions (FP), and false negatives (FN). In the issue detection task, TP is the screenshot correctly predicted as buggy; FN is the screenshot of the incorrectly predicted as buggy; TN is the screenshot correctly predicted as normal; FP is the screenshot incorrectly predicted as normal.

\begin{table}[htb]
\vspace{0.1in}
\caption{Confusion matrix}
\vspace{-0.05in}
\label{tab:RQ1-bug-dec-p}
\centering
\footnotesize
\begin{tabular}{p{1.0cm}<{\centering}  p{2.0cm}<{\centering} | p{1.5cm}<{\centering} p{1.8cm}<{\centering}}
\hline
\multicolumn{2}{c|}{\multirow{2}{*}{\textbf{Confusion matrix}}} & \multicolumn{2}{c}{\textbf{Observed}}\cr\cline{3-4}
 &  & True fault & Not true fault  \\
\hline
\multirow{2}{*}{\textbf{Predicted}} & True fault & TP & FP \cr
 & Not true fault & FN & TN \\
\hline
\end{tabular}
\vspace{-0.05in}
\end{table}

Precision is the proportion of screenshots that are correctly predicted as having UI display issues among all screenshots predicted as buggy:

\begin{equation}
\mathit{precision = \frac{\mathit{TP}}{\mathit{TP + FP}}}   
\end{equation}

Recall is the proportion of screenshots that are correctly predicted as buggy among all screenshots that really have UI display issues.
\begin{equation}
\mathit{recall = \frac{\mathit{TP}}{\mathit{TP + FN}}}   
\end{equation}

F1-score (F-measure or F1) is the harmonic mean of precision and recall, which combines both of the two metrics above.
\begin{equation}
\mathit{F1-score = \frac{\mathit{2 \times precision \times recall}}{\mathit{precision + recall}}}   
\end{equation}

In Section \ref{sec_results_RQ2}, in order to evaluate the issues localization performance of our proposed approach in RQ2, we employ two evaluation metrics, i.e., average precision (AP) and average recall (AR), which are commonly-used in object detection~\cite{ren2015faster}.
The AP and AR can more accurately and rigorously describe the localization performance of the {\tool}.
For all metrics, the higher the value, the better the performance. 
Among them, AP and AR are similar to precision and recall in image classification, but the evaluation content is different. First, select the prediction box whose confidence score is greater than 0.5~\cite{ren2015faster}, and calculate the ratio $IoU$ (intersection over union) of the intersection and union of the prediction box and ground truth box. The calculation approach is as follows: $IoU$ = the intersection of the predicted buggy region and the real buggy region / the union of the predicted buggy region and the real buggy region. The metric $IoU$ can solve the coverage problem. Then true positives (TP) is the number of detection boxes with $IoU \ge 0.5$. False positives (FP) is the number of detection boxes with $IoU \textless 0.5$, and the number of redundant detection boxes detected in the same ground truth box. False negatives (FN) is the number of ground truth box not detected. 

\section{Results and Analysis}
\label{sec_results}

\subsection{Issues Detection Performance (RQ1)}
\label{sec_results_RQ1}

We first present the issues detection performance of our proposed {\tool}, as well as the performance in terms of four categories of UI display issues in the data-generation dataset (Data-Gen) and the real-world dataset (Real-World) in Table \ref{tab:RQ1-bug-dec-p}.
In the data-generation dataset, with {\tool}, the average precision (P) is 0.843, indicating that an average of 84.3\% (837/993) of the screenshots which are predicted as having UI display issues are truly buggy. 
The average recall (R) is 0.837, indicating that an average of 83.7\% (837/1000) buggy screenshots can be found with {\tool}.
In the real-world dataset, with {\tool}, the average precision is 0.826, indicating that an average of 82.6\% (164/199) of the screenshots which are predicted as having UI display issues are truly buggy. 
The average recall is 0.821, indicating 82.1\% (164/200) buggy screenshots can be found with {\tool}.

\begin{table}[htb]
\caption{Issues detection performance (RQ1)}
\vspace{-0.05in}
\label{tab:RQ1-bug-dec-p}
\centering
\footnotesize
\begin{tabular}{p{2.8cm}<{\centering} | p{0.5cm}<{\centering} p{0.5cm}<{\centering} p{0.5cm}<{\centering} || p{0.5cm}<{\centering} p{0.5cm}<{\centering} p{0.5cm}<{\centering}}
\hline
\multirow{2}{*}{\textbf{Category}} &\multicolumn{3}{c||}{\textbf{Data-Gen}} &  \multicolumn{3}{c}{\textbf{Real-World}}\cr\cline{2-7}
 & \textbf{P} & \textbf{R} & \textbf{F1} & \textbf{P} & \textbf{R} & \textbf{F1} \\
\hline

Component occlusion & 0.768 & 0.751 & 0.759 & 0.750 & 0.735 & 0.742 \\ 

Text overlap & 0.851 & 0.813 & 0.832 & 0.810 & 0.790 & 0.800 \\ 

Missing image & 0.870 & 0.879 & 0.875 & 0.866 & 0.870 & 0.868 \\

NULL value & 0.881 & 0.904 & 0.892 & 0.878 & 0.890 & 0.883\\ 

\hline
\hline
 \textbf{Average} & \textbf{0.843} & \textbf{0.837} & \textbf{0.840} & \textbf{0.826} & \textbf{0.821} & \textbf{0.823}\\
\hline
\end{tabular}
\vspace{-0.05in}
\end{table}

Although our {\tool} is training on the data-generation dataset, the average precision and recall in the real-world dataset are only 0.017 and 0.016 lower than the data-generation dataset, which further shows the effectiveness of our {\tool}.

We then shift our focus to the top half of Table \ref{tab:RQ1-bug-dec-p}, i.e., the performance in terms of each category of UI display issues. 
All the four categories of UI display issues can be detected with a relative high precision and recall, i.e., the maximum precision and recall are 0.88 and 0.90, the mimimum precision and recall are 0.77 and 0.75 respectively,
The category \textit{null value} can be detected with the highest F1-score, indicating both precision (0.88) and recall (0.90) achieve a relatively high value. 
This might because screenshots with \textit{null value} bugs have relatively fixed pattern and the buggy area is relatively obvious, i.e., the screenshot as shown in Section \ref{sec_motivation}.
In comparison, the category \textit{component occlusion} is recognized with the lowest F1-score, e.g., 0.77 precision and 0.75 recall.
This is due to the fact that the pattern of this category is more diversified, and the buggy region is much smaller, i.e., the occlusion area of component only accounts for a mere of 10\% of the component area.

\begin{figure}[htb]
\centering
\vspace{0.2in}
\includegraphics[width=8.8cm]{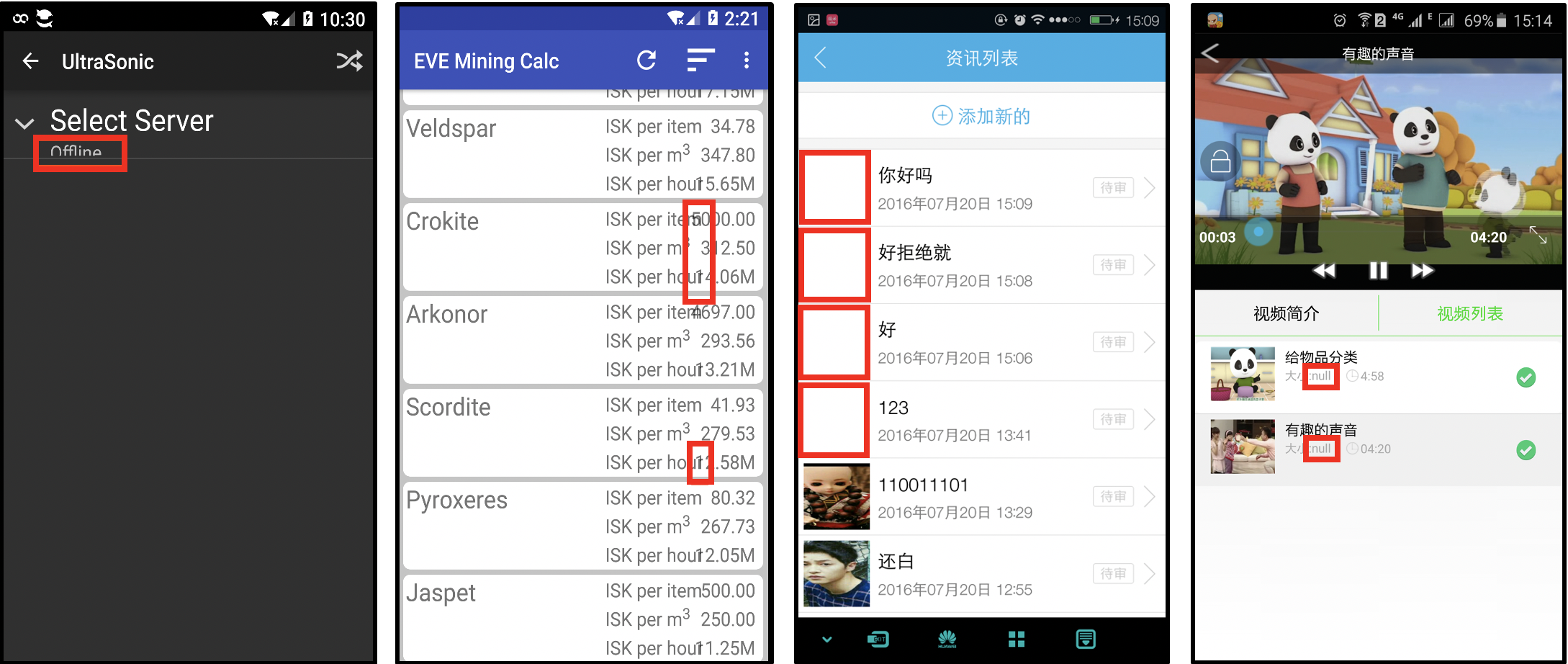}
\caption{Examples of bad case in issues detection (RQ1)}
\label{fig:detect-wrong}
\vspace{-0.05in}
\end{figure}

We further analyze the screenshots which are wrongly predicted as bug-free, with examples in Figure \ref{fig:detect-wrong}.
One common shared by these screenshots is that the buggy area is too tiny to be recognized even with human eye. 
Future work will focus more on improving the detection performance for these screenshots with attention mechanism and image magnification.

\subsubsection{\textbf{Performance Comparison with Baselines.}}
\label{sec_results_RQ1_1}

Table \ref{tab:RQ2-Baseline} shows the performance comparison with the baselines. 
We first compare the {\tool} with the 5 baselines in the top half of the table.
We can see that our proposed {\tool} is better than the 5 baselines, i.e., 23.2\% higher in precision and 17.1\% higher in recall compared with the best baseline (MLP).
This further indicates the effectiveness of {\tool}.
Besides, it also implies that {\tool} is especially good at hunting for the buggy screenshots from candidate ones.
MLP achieves the highest precision and recall among the baselines, indicating this deep learning approach is better at identifying the buggy screenshots.

\begin{table}[htb]
\caption{Performance comparison with baselines (RQ1)}
\vspace{-0.05in}
\label{tab:RQ2-Baseline}
\centering
\footnotesize
\begin{tabular}{p{2.5cm}<{\centering} | p{0.55cm}<{\centering} p{0.55cm}<{\centering} p{0.55cm}<{\centering} || p{0.55cm}<{\centering} p{0.55cm}<{\centering} p{0.55cm}<{\centering}}
\hline
\multirow{2}{*}{\textbf{Method}} &\multicolumn{3}{c||}{\textbf{Data-Gen}} &  \multicolumn{3}{c}{\textbf{Real-World}}\cr\cline{2-7}
 & \textbf{P} & \textbf{R} & \textbf{F1} & \textbf{P} & \textbf{R} & \textbf{F1} \\

\hline
SIFT-SVM & 0.525 & 0.518 & 0.521 & 0.500 & 0.495 & 0.497 \\ 
SIFT-KNN & 0.537 & 0.538 & 0.537 & 0.512 & 0.513 & 0.512 \\ 
SIFT-NB & 0.564 & 0.562 & 0.563 & 0.518 & 0.509 & 0.514 \\ 
SIFT-RF & 0.556 & 0.563 & 0.559 & 0.511 & 0.511 & 0.511 \\ 
\hline
SURF-SVM & 0.551 & 0.545 & 0.548 & 0.517 & 0.514 & 0.516 \\ 
SURF-KNN & 0.560 & 0.558 & 0.559 & 0.526 & 0.528 & 0.527 \\ 
SURF-NB & 0.586 & 0.590 & 0.588 & 0.543 & 0.535 & 0.539 \\ 
SURF-RF & 0.583 & 0.585 & 0.584 & 0.529 & 0.535 & 0.532 \\ 
\hline
ORB-SVM & 0.556 & 0.549 & 0.552 & 0.519 & 0.516 & 0.518 \\ 
ORB-KNN & 0.564 & 0.561 & 0.562 & 0.526 & 0.530 & 0.528 \\ 
ORB-NB & 0.586 & 0.590 & 0.588 & 0.547 & 0.540 & 0.544 \\ 
ORB-RF & 0.584 & 0.586 & 0.585 & 0.535 & 0.544 & 0.539 \\ 
\hline
MLP & 0.611 & 0.666 & 0.637 & 0.547 & 0.539 & 0.543 \\  
\hline
OwlEye & 0.790 & 0.778 & 0.784 & 0.764 & 0.743 & 0.753\\

\hline
\hline
\textbf{{\tool}} & \textbf{0.843} & \textbf{0.837} & \textbf{0.840} & \textbf{0.826} & \textbf{0.821} & \textbf{0.823}\\
\hline
\end{tabular}
\vspace{-0.05in}
\end{table}

We compare the performance of the newly-designed model and the old model in OwlEye with the same training and testing data, and results demonstrate that the newly-designed model in {\tool} achieves better performance, i.e., 0.843 vs. 0.790 in precision, and 0.837 vs. 0.778 in recall.
Our {\tool} has a obvious improvement in precision and recall, which may be due to the fact that the object detection task provides a more accurate bounding box for the UI display issues when training the model, which is more convenient for the model to learn the corresponding features. The training data of other baseline methods only have category labels. {\tool} performs well in issue detection, and does not need to label data manually, so it can better adapt to the different style of Android UI.

\subsection{Issue Localization Performance (RQ2)}
\label{sec_results_RQ2}

\begin{figure}[htb]
\vspace{0.2in}
\centering
\includegraphics[width=8.8cm]{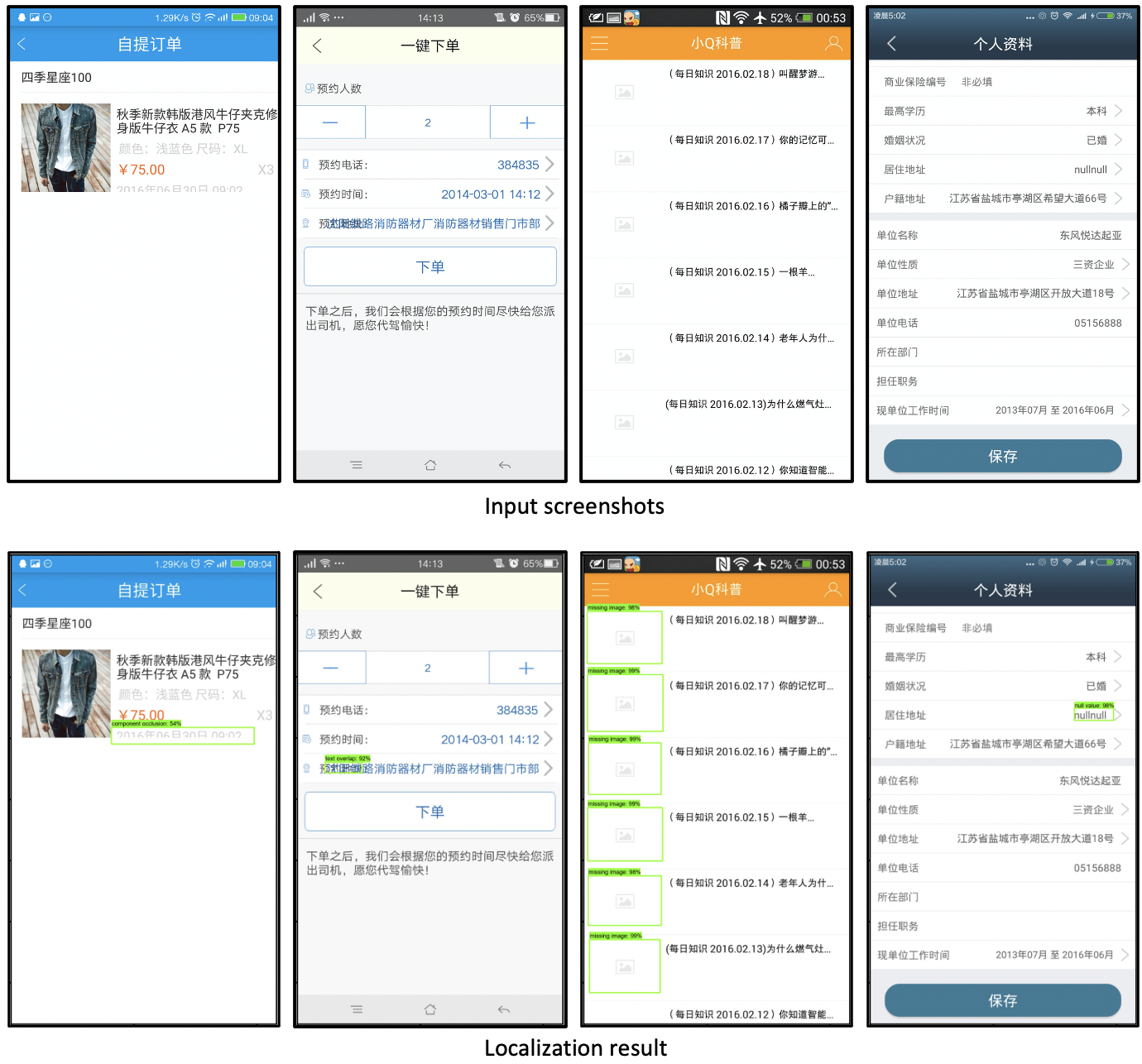}
\caption{Examples of issues localization (RQ2)}
\label{fig:localization}
\vspace{-0.05in}
\end{figure}

Figure \ref{fig:localization} presents the examples of our issues localization which highlights the buggy areas.
Since the localization result of OwlEye is in the form of heat map, in order to compare its performance with {\tool}, we use image binarization to determine the bounding box of the highlighted area of heat map, so that we can compare its performance with the newly proposed approach.

Table \ref{tab:RQ2-localization-result} shows the issues location performance of our proposed {\tool}. In the real-world dataset, the average AP(average precision) and AR(average recall) of {\tool} are 0.589 and 0.601 respectively. Due to the limitation of space, we only show the localization results in the real-world dataset, and the trend in the data-generation dataset is similar.

\begin{table}[htb]
\vspace{0.1in}
\caption{Issues localization performance (RQ2)}
\vspace{-0.05in}
\label{tab:RQ2-localization-result}
\centering
\footnotesize
\begin{tabular}{p{2.8cm}<{\centering} | p{1.0cm}<{\centering} p{1.0cm}<{\centering}|| p{1.0cm}<{\centering} p{1.0cm}<{\centering}}
\hline
\multirow{2}{*}{\textbf{Category}} &\multicolumn{2}{c||}{\textbf{OwlEye}} &  \multicolumn{2}{c}{\textbf{{\tool}}}\cr\cline{2-5}
 & \textbf{AP} & \textbf{AR}  & \textbf{AP} & \textbf{AR} \\
\hline

Component occlusion & 0.011 & 0.017 & 0.503 & 0.547 \\ 

Text overlap & 0.014 & 0.012 & 0.538 & 0.567 \\ 

Missing image & 0.103 & 0.121 & 0.773 & 0.765 \\

NULL value & 0.024 & 0.028 & 0.541 & 0.523 \\ 

\hline
\hline
\textbf{Average} & 0.038 & 0.045  & \textbf{0.589} & \textbf{0.601}\\
\hline
\end{tabular}
\vspace{-0.05in}
\end{table}

We then shift our focus to the top half of Table \ref{tab:RQ2-localization-result}, i.e., the issues location performance in terms of each category of UI display issues. All the four categories of UI display issues can be detected with a high AP and AR, i.e., in the real-world  dataset the mimimum AP and AR are 0.503 and 0.523 respectively. 
In the real-world  dataset, the category \textit{missing image} can be detected with the highest performance, indicating both AP (0.773) and AR (0.765) achieve a relatively high value. 
This might because screenshots with \textit{missing image} issues have large and relatively obvious buggy area.
In comparison, the category \textit{component occlusion} is recognized with the lowest performance, e.g., 0.503 AP and 0.547 AR.

\begin{figure}[htb]
\vspace{0.2in}
\centering
\includegraphics[width=8.8cm]{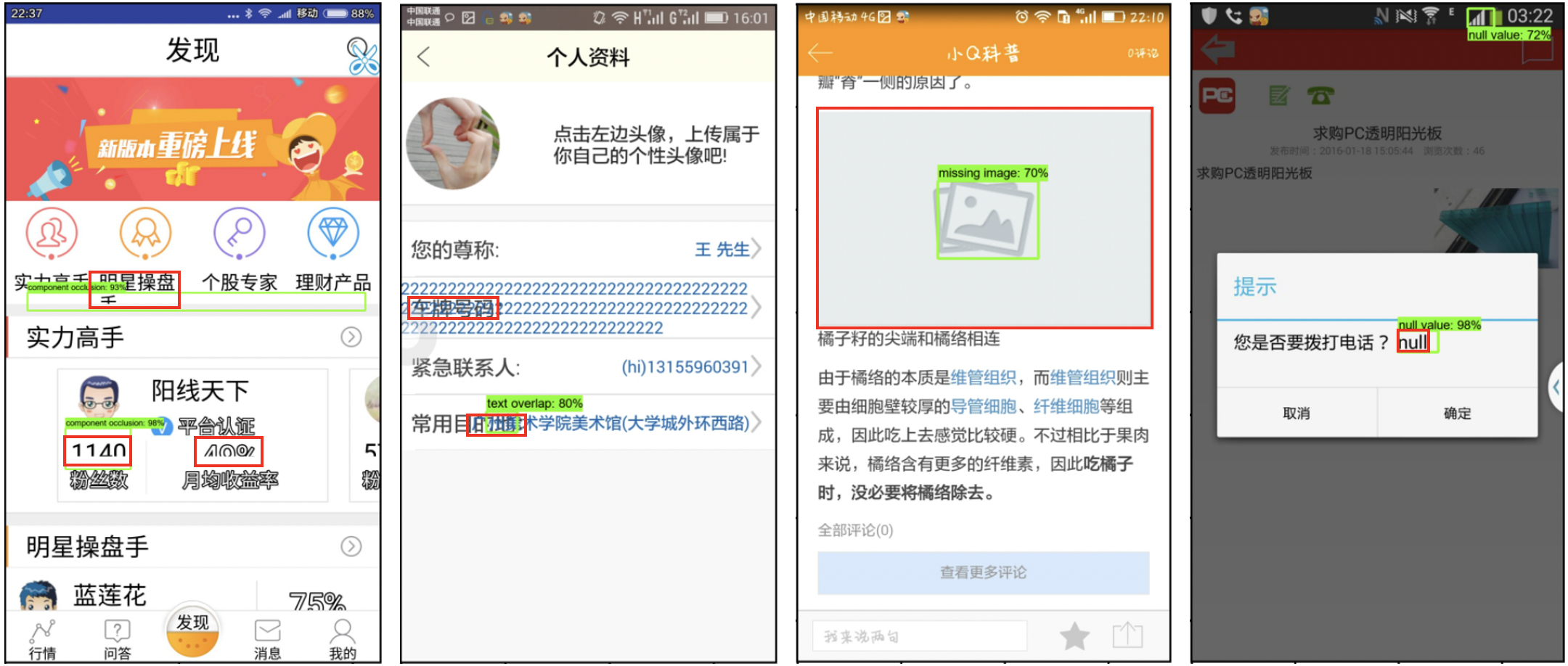}
\caption{Examples of bad case in issues localization (RQ2)}
\label{fig:localization-wrong}
\vspace{-0.05in}
\end{figure}

This is due to the fact that the buggy region is much smaller. As shown in Figure \ref{fig:localization-wrong}, the green one is the prediction bounding box, and the red one is the ground truth bounding box. The predicted bounding box area will be slightly larger than the real-world bounding box, resulting in IoU area less than 0.5, which will reduce the AP and AR. 
However, as shown in Figure \ref{fig:localization-wrong}, although the predicted localization result is judged to be wrong due to IoU less than 0.5, the indicated bug area is basically correct, which can also provide corresponding presentation for developers.

\begin{figure}[htb]
\centering
\includegraphics[width=8.8cm]{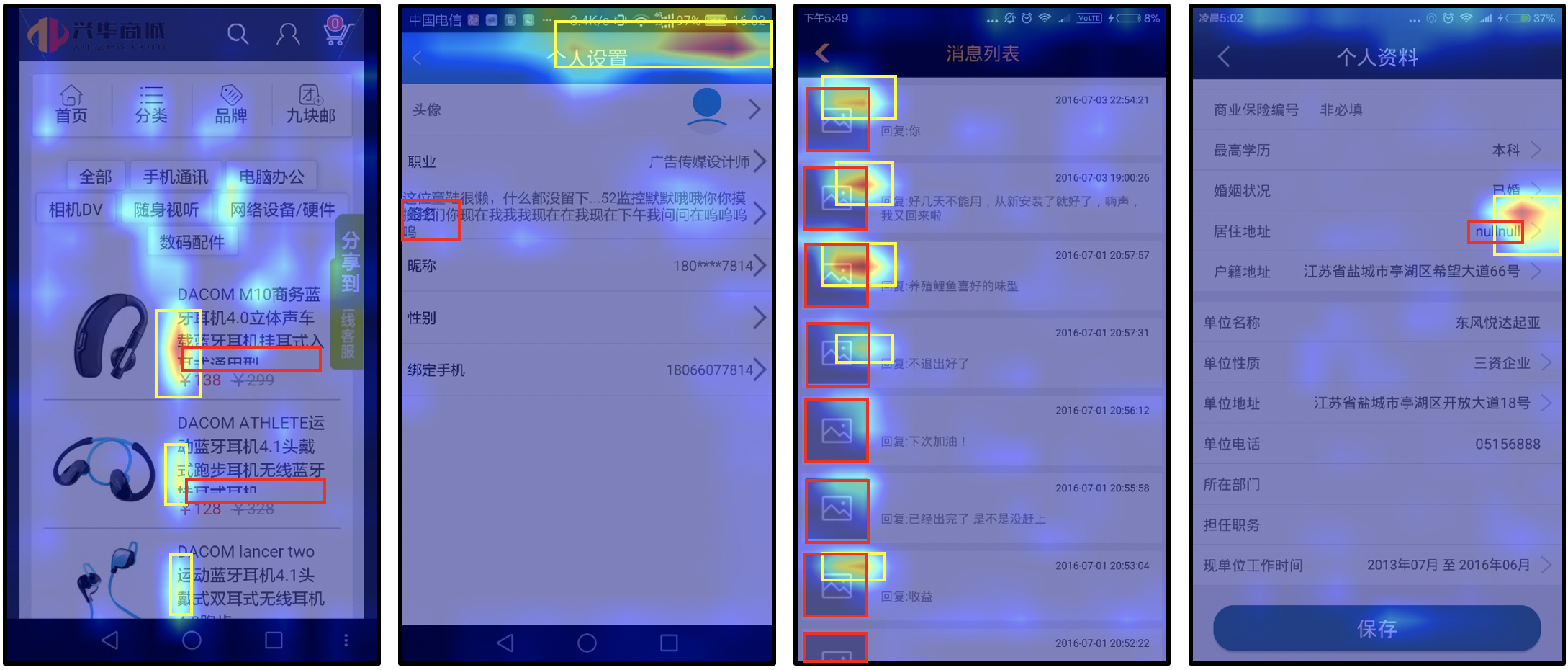}
\caption{Examples of bad case in OwlEye localization (RQ2)}
\label{fig:localization-wrong-owleye}
\vspace{-0.05in}
\end{figure}

Then we compare the issues localization performance of our {\tool} with OwlEye. As shown in Table \ref{tab:RQ2-localization-result} our {\tool} is obviously better than OwlEye, i.e., 55\% higher in AP and 56\% higher in AR compared with the OwlEye. We further analyze the bad case of issues localization with OwlEye. As shown in Figure \ref{fig:localization-wrong-owleye}, the yellow one is the prediction bounding box from OwlEye, and the red one is the ground truth bounding box. The main reason is that the visualization area of OwlEye is too large, which often contains the ground truth bounding box, and the IoU area is less than 0.5.

\begin{figure}[htb]
\centering
\vspace{0.05in}
\includegraphics[width=8.4cm]{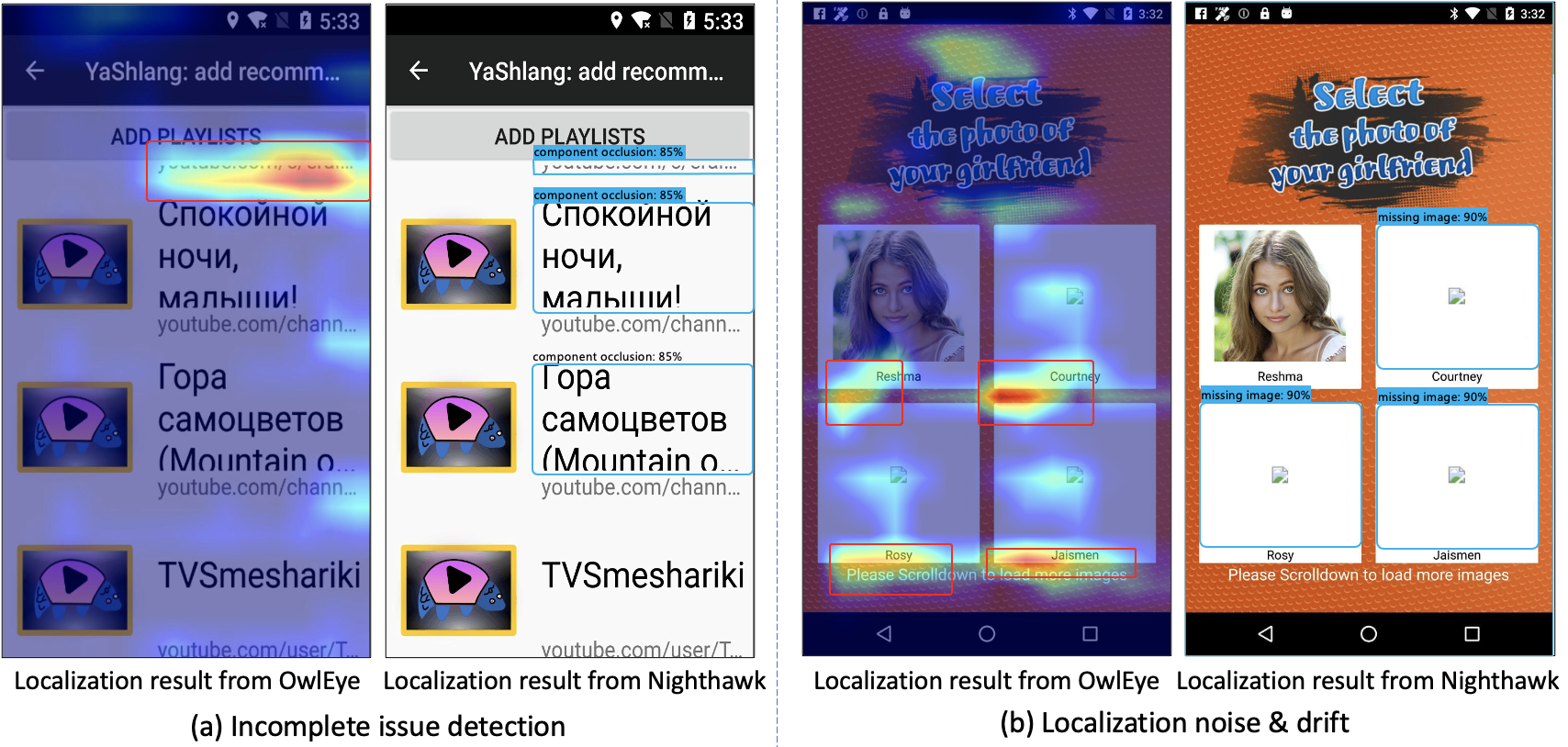}
\caption{Difference in localization of {\tool} and OwlEye}
\label{fig:different}
\vspace{-0.05in}
\end{figure}

Please note that the worse performance of OwlEye in AP and AR is due to the different evaluation criterion.
In the previous paper, we used manual evaluation to determine the accuracy of UI display issue localization, in which the participants are required to evaluate whether the localized area by OwlEye has overlap with the actual issue area.
By comparison, the AP and AR indicators require the IoU area to be more than 0.5, i.e., the highlighted issue area should be at least 50\% in common with the actual issue area. 
We observe three cases in which the human evaluation comes out with a high performance, yet the AP/AR suggests a low localization accuracy.
The first is incomplete issue detection as shown in Figure \ref{fig:different}(a). We can see that there are three issue areas, while OwlEye only highlights one of them. By comparison, the newly-proposed {\tool} can detect all of them, and achieve better AP/AR than OwlEye.
The second is localization noise as shown in Figure \ref{fig:different}(b).
OwlEye wrongly highlights another area in the lower left part of the screenshot, and would obtain lower AP/AR than {\tool}.
The third is the localization drift as shown in Figure \ref{fig:different}(b).
The newly-proposed {\tool} can perfectly localize the issue area, i.e., the whole image region, while the highlighted area by OwlEye does not hit the target accurately.

\subsection{\textbf{Contribution of Data Auto-generation (RQ3)}}
\label{sec_results_RQ3}

\begin{figure}[htb]
\centering
\includegraphics[width=8.8cm]{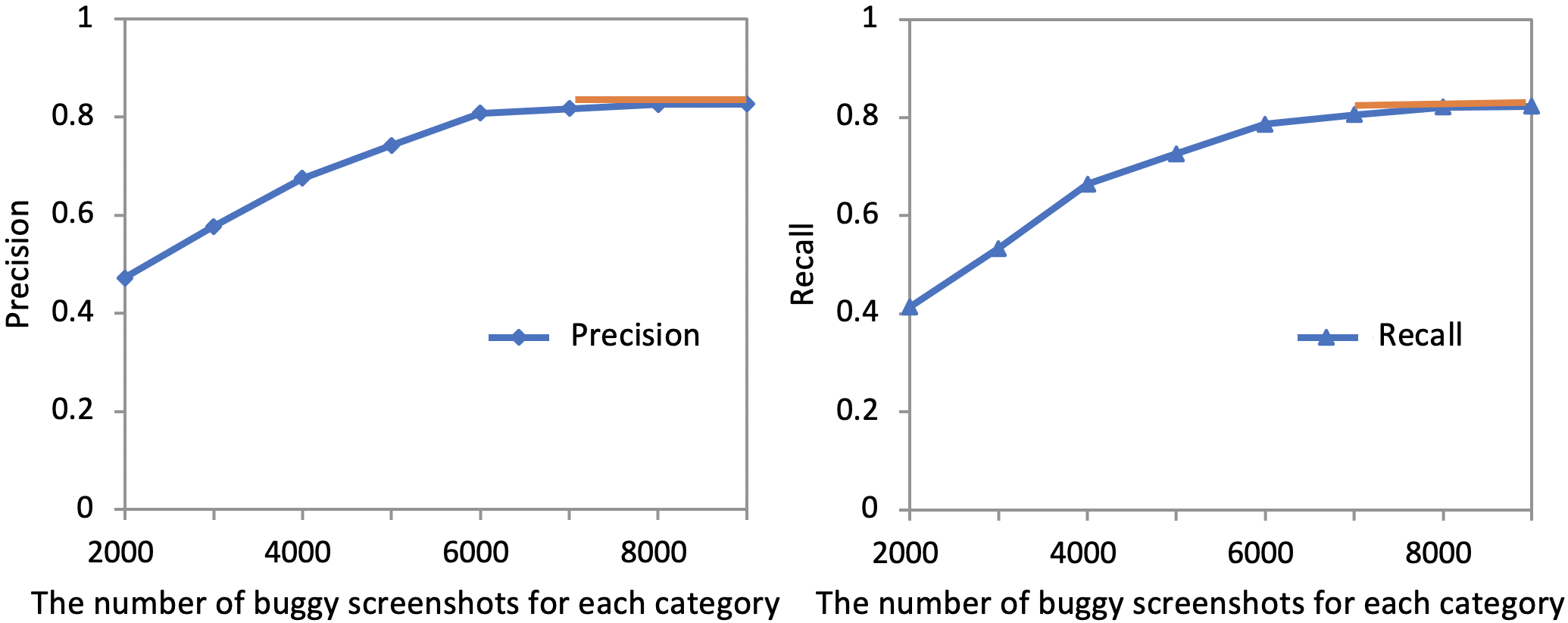}
\caption{Result of different training data configurations (RQ3)}
\label{fig:number data}
\vspace{-0.05in}
\end{figure}

We also conduct experiments to compare the issue detection performance of our {\tool} using different amounts of training data. The testing set is manually labeled data from the crowdtesting dataset, with 400 screenshots of each category (half positive and half negative samples, see Section \ref{subsec_experiment_dataset}). As shown in Figure \ref{fig:number data}, we randomly extract 2000-9000 screenshots with bugs as positive samples in the training set, and extract the same number of bug-free screenshots as negative samples according to the same settings and proportions to form the training set.

Figure \ref{fig:number data} shows the performance of UI display issues detection in terms of different amounts of training data, i.e., the number of negative samples is from 2000 to 9000.

We can see that both precision and recall improve with the increase of data volume, indicating the value of data auto-generation for effective UI display issues detection. 
Specifically, compared with the training data of 2000 buggy screenshots, the improvement of 35\% and 41\% in average precision and recall of 8000 buggy screenshot are observed respectively. 
The larger improvement in recall and precision indicates that, with more data, more screenshots with UI display issues can be found.
This might because the model parameters of the object detection task are more than the image classification model, thus more training samples are needed to train the model.

As shown in Figure \ref{fig:number data}, with the increase of the amount of data, the improvement of model performance is gradually decreasing. For example, the effect of 8000 buggy screenshots and 9000 buggy screenshots almost remains unchanged. At the same time, too much data will increase the cost of training model, so we chose 8000 buggy screenshots as our final training data.

\subsubsection{\textbf{Data Auto-generation Performance}}
\label{sec_results_RQ3_1}

We investigate the contribution of data auto-generation by comparing the issues detection performance of the data auto-generation method in {\tool}(Auto-DataGen) with that of the data augmentation method in OwlEye (DataAug) (details are in Section \ref{subsec_experiment_dataset}).
We used the same amount of generated data to train our {\tool} to complete the experiment.

\begin{table}[H]
\vspace{0.1in}
\caption{Contribution of data auto-generation (RQ4)}
\vspace{-0.05in}
\label{tab:RQ4-data-aug-2.0}
\centering
\footnotesize
\begin{tabular}{p{2.8cm}<{\centering} | p{0.45cm}<{\centering} p{0.45cm}<{\centering} p{0.45cm}<{\centering} || p{0.45cm}<{\centering} p{0.45cm}<{\centering} p{0.45cm}<{\centering}}
\hline
\multirow{2}{*}{\textbf{Category}} &\multicolumn{3}{c||}{\textbf{DataAug}} &  \multicolumn{3}{c}{\textbf{Auto-DataGen}}\cr\cline{2-7}
 & \textbf{P} & \textbf{R} & \textbf{F1} & \textbf{P} & \textbf{R} & \textbf{F1} \\
\hline

Component occlusion & 0.672 & 0.605 & 0.636 & 0.750 & 0.735 & 0.742 \\ 

Text overlap & 0.716 & 0.645 & 0.679 & 0.810 & 0.790 & 0.800 \\ 

Missing image & 0.785 & 0.765 & 0.774 & 0.866 & 0.870 & 0.868 \\

NULL value & 0.711 & 0.795 & 0.803 & 0.878 & 0.890 & 0.883\\ 

\hline
\hline
\textbf{Average} & 0.746 & 0.703 & 0.723 & \textbf{0.826} & \textbf{0.821} & \textbf{0.823}\\
\hline

\end{tabular}
\vspace{-0.05in}
\end{table}

As shown in Table \ref{tab:RQ4-data-aug-2.0}, results show that, 8\% and 12\% improvement are observed respectively for average precision and recall. Specifically, issues detection performance in \textit{component occlusion} category undergoes the largest improvement in F1-score.
This might because the data auto-generation method (DataGen({\tool})) pays more attention to the diversity of training screenshots and the generation of tiny region bugs (details are in Section \ref{subsec_approach_augmentation}). After using data auto-generation method (DataGen({\tool})), the diversity of the training screenshots significantly improves the performance.

\begin{table}[htb]
\vspace{0.1in}
\caption{Confirmed or fixed issues (RQ4)}
\vspace{-0.05in}
\label{tab:RQ3-Usefulness}
\centering
\footnotesize
\begin{tabular}{p{2.5cm}<{\centering} | p{1.1cm}<{\centering} | p{1.3cm}<{\centering} | p{0.9cm}<{\centering} | p{0.9cm}<{\centering}}
\hline
\textbf{APP Name} & \textbf{Category} & \textbf{Download} & \textbf{IssueId} & \textbf{Status}\\
\hline
\multicolumn{5}{c}{\textbf{Apps From Google Play}}\\
\hline
SHAREit & Tools & 500M+ & email & fixed\\

ShareMe & Tools & 500M+ & email & fixed\\

Perfect Piano & Music & 50M+ & email & confirm\\

Music Player & Music & 50M+ & email & confirm\\  

Status Saver & Product & 50M+ & email & fixed\\  

Nimo TV & Enter & 50M+ & email & fixed\\  

Nox security & Tool & 10M+ & email & fixed\\  

DegooCloud & Tool & 10M+ & email & fixed\\  

Proxynel & Tool & 10M+ & email & confirm\\  

Secure VPN & Tool & 10M+ & email & confirm\\  

Thunder VPN & Tool & 10M+ & email & fixed \\

Sweatcoin & Health & 10M+ & email & fixed \\

ApowerMirror & Tool & 5M+ & email & confirm\\  

PUB Gfx & Libraries & 5M+ & email & fixed\\  

MediaFire & Product & 5M+ & email & confirm\\  

Paytm & Finance & 1M+ & email & confirm \\

Playnimes Animes & Video & 1M+ & email & fixed\\  

Postegro & Commun & 500K+ & email & fixed\\   

Deezer Player & Music & 500K+ & email & fixed\\  

Air China & Travel & 100K+ & email & fixed \\
\hline
\hline
\multicolumn{5}{c}{\textbf{Apps From F-droid}}\\
\hline
Librera Reader & Reading & 10M+ & \#652 & fixed \\ 

AdGuard & Secur & 5M+ & \#243 & confirm\\

Kiwix & Books & 1M+ & \#2566 & confirm\\

MTG Familiar & Utilities & 500K+ & \#512 & fixed \\  

My Expenses & Finance & 500K+ & \#715 & fixed \\ 

WiGL & Connect & 500K+ & \#460 & fixed \\

Linphone & Commun & 500K+ & \#965 & confirm \\ 

Transdroid & Tool & 100K+ & \#542 & confirm \\ 

Tutanota & Commun & 100K+ & \#2527 & confirm \\

Onkyo & Music & 10K+ & \#138 & fixed \\

Aria2App & Inter & 10K+ & \#140 & confirm \\

NewPipeLegacy & Media & 8K+ & \#24 & fixed \\ 

Noice & Media & 10K+ & \#458& fixed \\ 

EasyRepost & Media & 10K+ & \#9 & confirm \\

Ahorcado & Game & 10K+ & \#3 & fixed \\

LessPass & Product & 5K+ & \#519 & fixed \\

SimpleLogin & Internet & 5K+ & \#11 & fixed \\

DSAAssistent & Game & 1K+ & \#12 & confirm \\

Dagger & Enter & 1K+ & \#12 & fixed \\

BetterUntis & Edu & 1K+ & \#168 & confirm \\

ProExpense & Internet & 1K+ & \#23 & fixed \\

TrailSense & Navig & 500+ & \#338 & confirm \\

CEToolbox & Medical & 500+ & \#4 & confirm \\ 

AppIntro & Image & 500+ & \#915 & fixed \\

logger & Connect & 500+ & \#92 & fixed\\

Hauk & Navig & 300+ & \#162 & confirm \\

Shosetsu & Reading & 100+ & \#91 & fixed \\

Memory & Edu & 100+ & \#5 & fixed \\

DailyFinance & Finance & 50+ & \#23 & fixed \\

LeMondeRssr & Tool & 50+ & \#34 & confirm\\

Weather & Tool & 50+ & \#1 & fixed \\  

OpenTracks-OSM  & Health & 50+ & \#26 & fixed \\  

Yucata Envoy & Tool & 50+ & \#3 & confirm \\  

ClassyShark3 & Tool & 50+ & \#3 & confirm \\  

VlcFreemote & Media & 50+ & \#24 & confirm \\

\hline
\end{tabular}
\vspace{-0.05in}
\end{table}

\subsection{Usefulness Evaluation (RQ4)}
\label{sec_results_RQ4}

To further assess the usefulness of our {\tool}, we randomly sample 2,000 Android applications from F-Droid\footnote{http://f-droid.org/} and 2,000 Android applications from Google play\footnote{http://play.google.com/store/apps}, including many new apps released on 2019 and 2020.
Note that none of these apps appear in our training dataset.

We use DroidBot, which is a commonly-used lightweight Android test input generator~\cite{DroidBot}, for exploring the mobile apps and take the screenshot of each UI pages. 
Among the 4,000 collected apps, 70\% (2785/4000) apps can be successfully run with Droidbot, and only 33\% (1328/4000) of the apps can be fetched with more than one screenshot, as they require register or authenticate to explore more screenshots which cannot be done by DroidBot.
For the remaining 1328 apps, an average of eight screenshots are obtained for each app. 
We then feed those screenshots to {\tool} for detecting if there are any UI display issues.
Once a display issue is spotted, we create a bug report by describing the issue attached with buggy UI screenshot.
Finally, we report them to the app development team through issue reports or emails.

\begin{figure}[htb]
\centering
\vspace{0.1in}
\includegraphics[width=8.8cm]{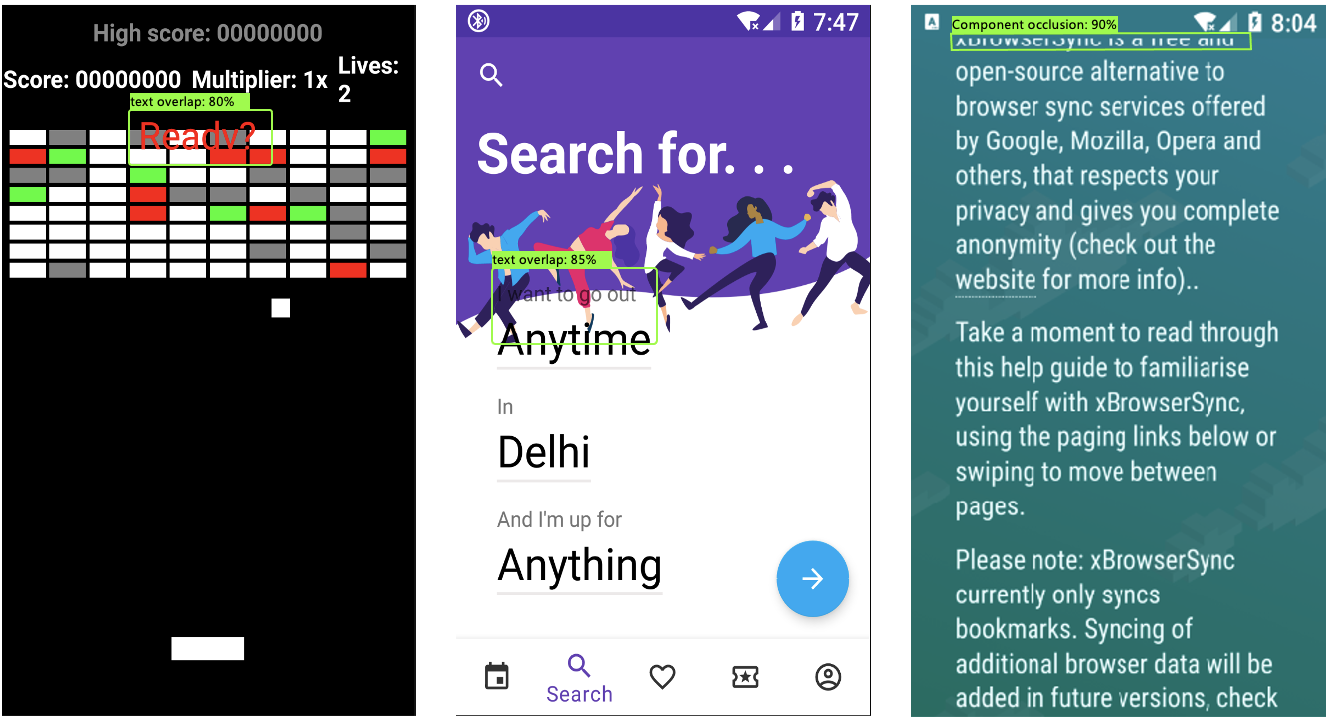}
\caption{UI display issues detected by Nighthawk (RQ4)}
\label{fig:usefull-d}
\vspace{-0.05in}
\end{figure}

Table \ref{tab:RQ3-Usefulness} shows all bugs spotted by our {\tool}, and more detailed information of detected bugs can be seen in our website\textsuperscript{\ref{github}}.
For F-Droid applications, 82 UI display issues are detected, among which 23 have been fixed and another 18 have been confirmed by the developers. 
For Google Play, 69 UI display issues are detected, among which 25 have been fixed and another 9 have been confirmed by the developers.
These fixed or confirmed bug reports further demonstrate the effectiveness and usefulness of our proposed approach in detecting UI display issues. 

The results reveal that {\tool} can not only cover all the issues detected by OwlEye, but also find 66 more issues than OwlEye.
The precision of {\tool} is 81\% (151/186), which is 14\%((81\%-71\%)/71\%) higher than that of OwlEye.
As shown in the Figure \ref{fig:usefull-d}, the {\tool} can detect more issues with small buggy area than OwlEye.
In addition, when developers confirm the issue report submitted by {\tool}, they say that the UI display issue has a great impact on the user experience and needs to be repaired as soon as possible. 
These replies also prove the necessity of {\tool} to detect UI display issue.
\section{Discussion}
\label{sec_discussion}

\textbf{\textit{Generality across platforms.}} 
Almost all the existing studies of GUI bug detection~\cite{zein2016systematic, lamsa2017comparison,DroidBot} are designed for a specific platform, e.g., Android, which limits its applicability in real-world practice.
In comparison, the primary idea of our proposed {\tool} is to detect UI display issues from the screenshots generated when running the applications.
Since the screenshots from different platforms (e.g., Android, iOS, Mac OS and Windows) exert almost no difference, our approach can be generalized for UI display issues detection in other platforms.

\begin{figure}[htb]
\centering
\includegraphics[width=8.8cm]{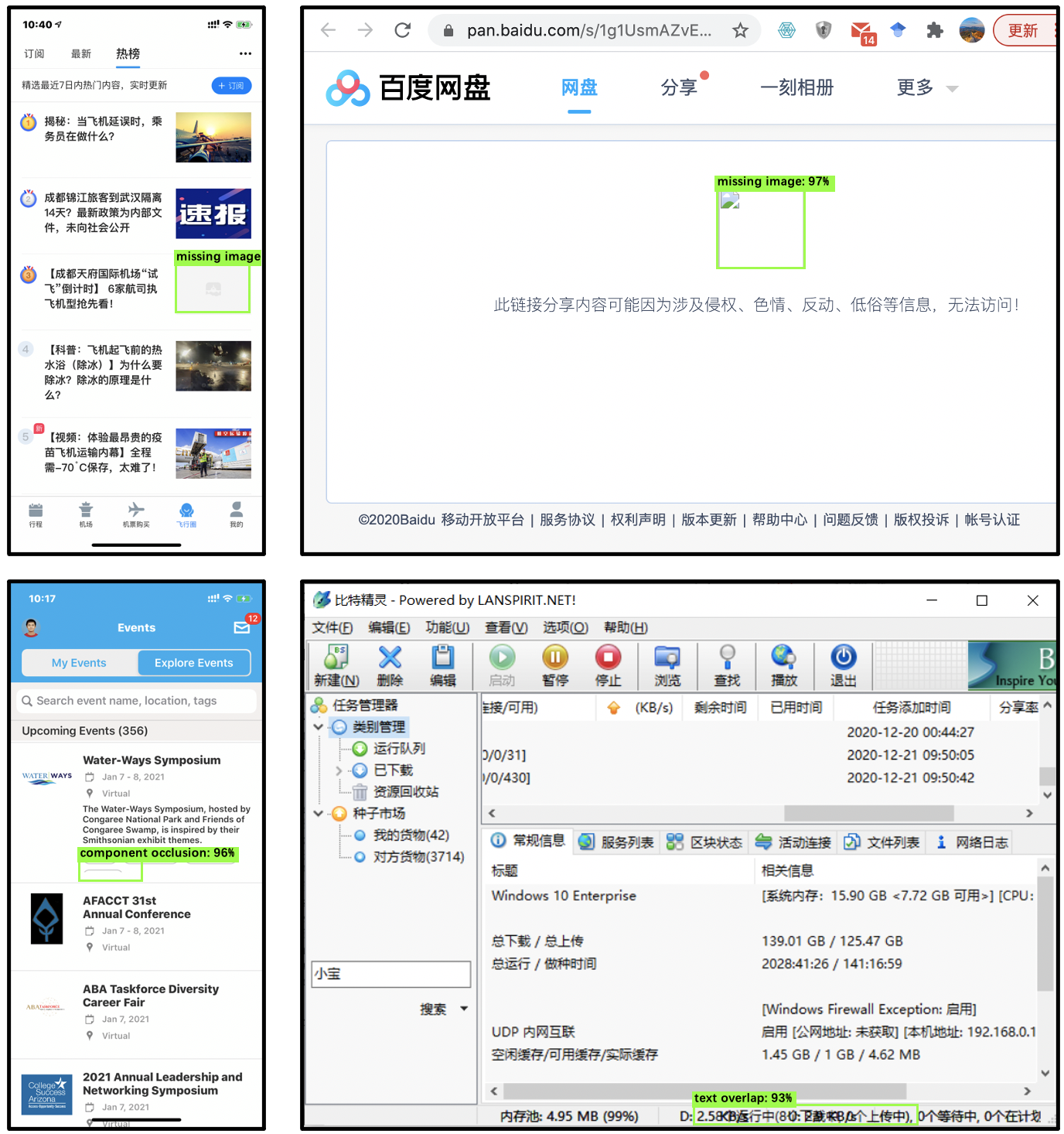}
\caption{Examples of different platforms}
\label{fig:different platform}
\vspace{-0.05in}
\end{figure}

As shown in Figure \ref{fig:different platform}, we have conducted a small scale experiment for three other popular platform, i.e., iOS, Mac OS and Windows, and experiment on 80 screenshots with UI display issues collected in our daily-used applications.
Results show that our proposed {\tool} can accurately detect 86.3\% (69/80) of the buggy screenshots.
This further demonstrates the generality of {\tool}, and we will conduct more thorough experiment in future.

\textbf{\textit{Generality across languages.}} 
Another advantage of {\tool} is that it can be applied for UI display issues detection in terms of different display languages of the application.
The testing data of the experiment for RQ1 contains the screenshots in Chinese, while the experiment for RQ3 relates with the screenshots in English, which demonstrates the generality of our approach across languages. 

\begin{figure}[h]
\centering
\includegraphics[width=8.8cm]{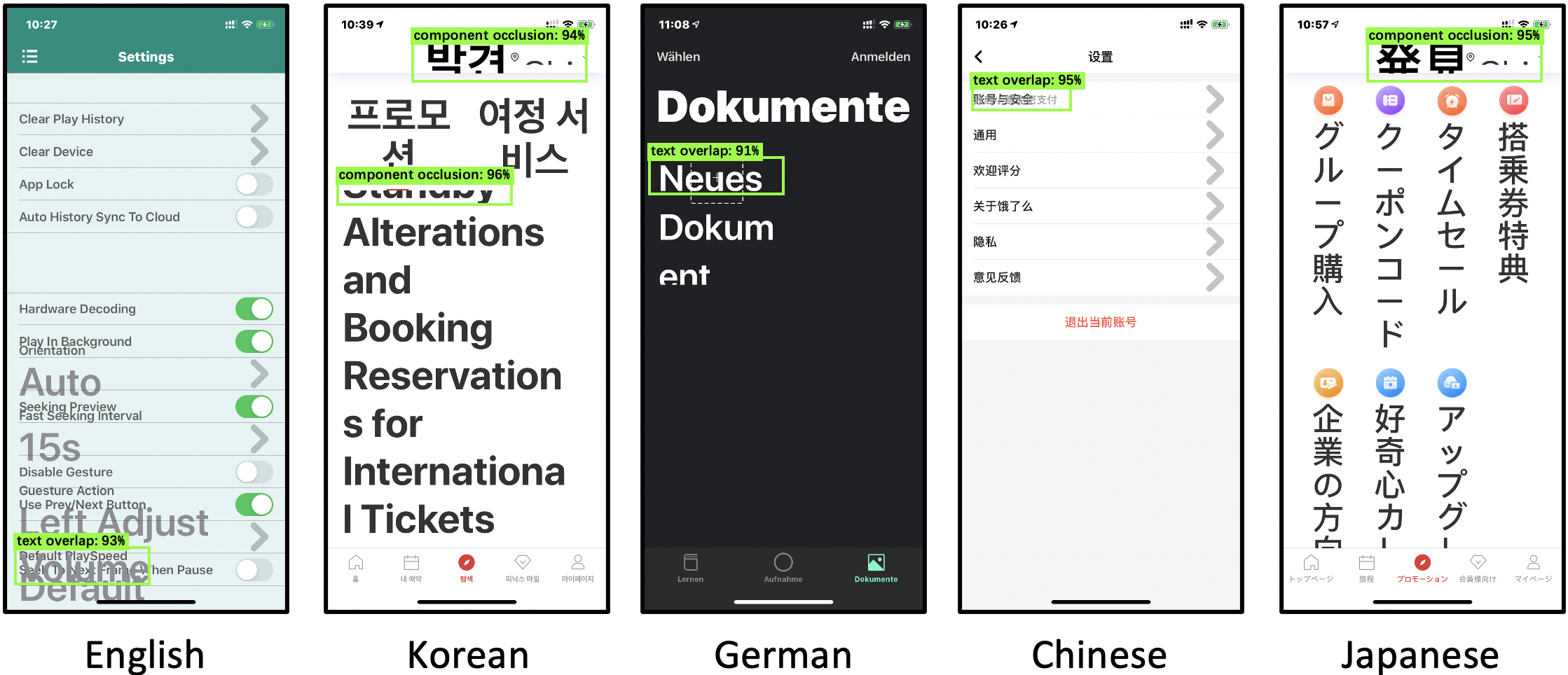}
\caption{Examples of different language settings on iPhone}
\label{fig:different language}
\vspace{-0.05in}
\end{figure}

As shown in Figure \ref{fig:different language}, we also collect 80 screenshots with UI display issues in three other languages (i.e. German, Japanese and Korean) from the applications in RQ3, and run our approach for bug detection. 
Results show that our proposed {\tool} can accurately detect 87.5\% (70/80) of the buggy screenshots, which further demonstrates the feasibility of {\tool}.

\textbf{\textit{Potential with more effective automatic testing tool.}} 
Results in RQ3 have demonstrated the usefulness of {\tool} in real-world practice being integrated with automatic testing tool as DroidBot.
However, we have mentioned in Section \ref{sec_results_RQ3} that some applications can not be run with DroidBot, and some can only be fetched with one screenshot due to the shortcoming of DroidBot, both of which limit the full exploration of screenshots. 
If armed with a more effective automatic testing tool, {\tool} should play a bigger role in detecting UI display issues in real-world practice.

\vspace{-0.05in}
\section{Related Work}
\textbf{Mobile App GUI.}
GUI provides a visual bridge between applications and users, and the quality of UI design is also widely concerned.
Therefore, many researchers are working on assisting developers or designers in the GUI search~\cite{reiss2018seeking, behrang2018guifetch,chen2020wireframe,chen2019gallery,zhao2019actionnet,yang2021uis,yang2021don,chen2021should} based on image features, GUI code generation~\cite{nguyen2015reverse, chen2018ui, moran2018machine,chen2019storydroid,chen2019gui,feng2021auto,zhao2021guigan,zhao2020seenomaly} based on computer vision techniques. Chen et al.~\cite{Chen2020From} introduce a novel approach based on a deep neural network for encoding both the visual and textual information to recover the missing tags for existing UI examples so that they can be more easily found by text queries.
Chen et al.~\cite{Chen2020Object} studied the limitations and effective design of object detection method based on deep learning in detecting UI components. In addition, they designed a new top-down strategy from coarse to fine, and combined it with mature GUI text deep learning model.
Moran et al.~\cite{moran2018automated} check if the implemented GUI violates the original UI design by comparing the images similarity with computer vision techniques.
A follow-up work by them~\cite{moran2018detecting} further detects and summarizes GUI changes in evolving mobile applications. 
Different from these works, our works are focusing on detecting the GUI display issues to help improve the app quality.

\textbf{Automated GUI Testing. }
To ensure that GUI is working well, there are many static linting tools to flag programming errors, bugs, stylistic errors, and suspicious constructs~\cite{zhao2020seenomaly,chen2020unblind}.
For example, Android Lint~\cite{lint} reports over 260 different types of Android bugs, including correctness, performance, security, usability and accessibility~\cite{chen2021accessible}.
StyleLint~\cite{stylelint} helps developers avoid errors and enforce conventions in styles.
Different from static linting, automatic GUI testing~\cite{mirzaei2016reducing, baek2016automated, su2017guided} dynamically explores GUIs of an app.
Several surveys~\cite{zein2016systematic, lamsa2017comparison} compare different tools for GUI testing for Android apps.
Some testing works focus on more specific UI issues such as UI rendering delays and image loading.
Gao et al.~\cite{gao2017every} and Li et al.~\cite{li2019characterizing} analyzed the possible problems in UI rendering, and developed automatic approaches to detect them.
Nayebi et al.~\cite{nayebi2012state} and Holzinger et al.~\cite{holzinger2012making} found that different resolutions of the mobile devices have brought challenges in Android app design and implementation.
Recently, deep learning based techniques~\cite{white2019improving, degott2019learning} have been proposed for automatic GUI testing.
Unlike traditional GUI testing which explores the GUIs by dynamic program analysis, they~\cite{white2019improving, degott2019learning} use computer vision techniques to detect GUI components on the screen to determine next actions.

The above mentioned GUI testing techniques focus on functional testing, while our work is more about non-functional testing i.e., GUI visual issues which will not cause app crash, but negatively influence the app usability.
The UI display bugs detected by our approach are mainly caused by the app compatibility~\cite{zhang2015compatibility, ki2019mimic} due to the different devices and Android versions. It is highly expensive and extremely difficult for the developers covering all the popular contexts when conducting testing.
Besides, different from these works based on static or dynamic code analysis, our work only requires the screenshot as the input.
Such characteristic enables our light-weight computer vision based method, and also makes our approach generalised to any platform including Android, IOS, or IoT devices.

\textbf{Web App Display Issues Detection. }
Because web apps also run on devices with a variety of viewport widths, there are some similarities between the mobile UI display issues detection technique in this paper and those testing approaches of web mobile apps. 
Kondratova and Goldfarb~\cite{kondratova2011culturally} explored the UI design preferences of different countries and cultures around the world.
WebSee~\cite{mahajan2015detection,mahajan2015websee} and FieryEye~\cite{mahajan2016using} utilized computer vision and image processing techniques to compare a browser rendered test webpage with an oracle image to find visual differences. 
BROWSERBITE~\cite{saar2016browserbite} combined image comparison methods with machine learning techniques to detect cross-browser differences in a web page.
Mahajan et al.~\cite{DBLP:conf/icse/MahajanAMH18, DBLP:conf/apsec/MahajanGPH16} studied the automatic detection and repair of mobile friendly problems in web pages. 
The XFix~\cite{DBLP:conf/issta/MahajanAMH17a} repaired layout Cross Browser Issues (XBIs) presentation failures arising from the inconsistencies in the rendering of a website across different browsers.
Additionally, IFIXCC~\cite{mahajan2021effective} repaired internationalization failures using search-based and clustering techniques, while CBRepair~\cite{alameer2019efficiently} addresses the same types of failures but using constraint solving.
Althomali et al.~\cite{DBLP:conf/icst/AlthomaliKM19} presented an automated approach that extracts the responsive layout of two versions of a page and compares them, alerting developers to the differences in layout that they may wish to investigate further.
Different from web apps, Android apps have more types of UI display issues (\textit{component occlusion, text overlap, missing image, null value, blurred screen}) and smaller issue area.
Therefore, our work adopts the method of computer vision for UI display issue detection, and uses the data generation method to improve the performance and robustness of the {\tool}. 
\section{Conclusion}
\label{sec_conclusion}

Improving the quality of mobile applications, especially in a proactive way, is of great value and always encouraged. 
This paper focuses on automatic detecting the UI display issues from the screenshots generated during automatic testing. 
The proposed {\tool} is proven to be effective in real-world practice, i.e., 75 confirmed or fixed previously-undetected UI display issues from popular Android apps.
{\tool} also achieves more than 17\% and 23\% boost in recall and precision compared with the best baseline.
As the first work of its kind, we also contribute to a systematical investigation of UI display issues in real-world mobile apps, as well as a large-scale dataset of app UIs with display issues for follow-up studies.

In the future, we will keep improving our model for better performance in the classification.
Apart from the display issue detection, we will further locate the root cause of these issues in our future work.
Then we will develop a set of tools for recommending patches to developers for fixing display bugs.

\ifCLASSOPTIONcompsoc
  \section*{Acknowledgments}
\else
  \section*{Acknowledgment}
\fi

This work is supported by the National Key Research and Development Program of China under grant No.2018YFB1403400, National Natural Science Foundation
of China under Grant No. 62072442, No. 62002348, and Youth Innovation Promotion Association Chinese Academy of Sciences.

\ifCLASSOPTIONcaptionsoff
  \newpage
\fi

\bibliographystyle{IEEEtran}

\bibliography{reference}

\begin{thebibliography}{10}
\providecommand{\url}[1]{#1}
\csname url@samestyle\endcsname
\providecommand{\newblock}{\relax}
\providecommand{\bibinfo}[2]{#2}
\providecommand{\BIBentrySTDinterwordspacing}{\spaceskip=0pt\relax}
\providecommand{\BIBentryALTinterwordstretchfactor}{4}
\providecommand{\BIBentryALTinterwordspacing}{\spaceskip=\fontdimen2\font plus
\BIBentryALTinterwordstretchfactor\fontdimen3\font minus
  \fontdimen4\font\relax}
\providecommand{\BIBforeignlanguage}[2]{{%
\expandafter\ifx\csname l@#1\endcsname\relax
\typeout{** WARNING: IEEEtran.bst: No hyphenation pattern has been}%
\typeout{** loaded for the language `#1'. Using the pattern for}%
\typeout{** the default language instead.}%
\else
\language=\csname l@#1\endcsname
\fi
#2}}
\providecommand{\BIBdecl}{\relax}
\BIBdecl

\bibitem{jansen1998graphical}
\BIBentryALTinterwordspacing
B.~J. Jansen, ``The graphical user interface,'' \emph{SIGCHI Bull.}, vol.~30,
  no.~2, p. 22–26, Apr. 1998. [Online]. Available:
  \url{https://doi.org/10.1145/279044.279051}
\BIBentrySTDinterwordspacing

\bibitem{liu2022Guided}
\BIBentryALTinterwordspacing
Z.~Liu, C.~Chen, J.~Wang, and Q.~Wang, ``Guided bug crush: Assist manual gui
  testing of android apps via hint moves,'' in \emph{CHI 2022}, 2022. [Online].
  Available: \url{https://doi.org/10.1145/3491102.3501903}
\BIBentrySTDinterwordspacing

\bibitem{DBLP:conf/sigsoft/SuLC0W21}
\BIBentryALTinterwordspacing
Y.~Su, Z.~Liu, C.~Chen, J.~Wang, and Q.~Wang, ``Owleyes-online: a fully
  automated platform for detecting and localizing {UI} display issues,'' in
  \emph{{ESEC/FSE} '21: 29th {ACM} Joint European Software Engineering
  Conference and Symposium on the Foundations of Software Engineering, Athens,
  Greece, August 23-28, 2021}.\hskip 1em plus 0.5em minus 0.4em\relax {ACM},
  2021, pp. 1500--1504. [Online]. Available:
  \url{https://doi.org/10.1145/3468264.3473109}
\BIBentrySTDinterwordspacing

\bibitem{wei2016taming}
L.~Wei, Y.~Liu, and S.-C. Cheung, ``Taming android fragmentation:
  Characterizing and detecting compatibility issues for android apps,'' in
  \emph{Proceedings of the 31st IEEE/ACM International Conference on Automated
  Software Engineering}, 2016, pp. 226--237.

\bibitem{RAD}
J.~Martin, \emph{Rapid Application Development}.\hskip 1em plus 0.5em minus
  0.4em\relax USA: Macmillan Publishing Co., Inc., 1991.

\bibitem{mirzaei2016reducing}
N.~Mirzaei, J.~Garcia, H.~Bagheri, A.~Sadeghi, and S.~Malek, ``Reducing
  combinatorics in gui testing of android applications,'' in \emph{Software
  Engineering (ICSE), 2016 IEEE/ACM 38th International Conference on}.\hskip
  1em plus 0.5em minus 0.4em\relax IEEE, 2016, pp. 559--570.

\bibitem{baek2016automated}
Y.-M. Baek and D.-H. Bae, ``Automated model-based android gui testing using
  multi-level gui comparison criteria,'' in \emph{Proceedings of the 31st
  IEEE/ACM International Conference on Automated Software Engineering}.\hskip
  1em plus 0.5em minus 0.4em\relax ACM, 2016, pp. 238--249.

\bibitem{su2017guided}
T.~Su, G.~Meng, Y.~Chen, K.~Wu, W.~Yang, Y.~Yao, G.~Pu, Y.~Liu, and Z.~Su,
  ``Guided, stochastic model-based gui testing of android apps,'' in
  \emph{Proceedings of the 2017 11th Joint Meeting on Foundations of Software
  Engineering}.\hskip 1em plus 0.5em minus 0.4em\relax ACM, 2017, pp. 245--256.

\bibitem{2017AutomatedGUI}
\BIBentryALTinterwordspacing
K.~Moran, M.~Linares-V\'{a}squez, and D.~Poshyvanyk, ``Automated gui testing of
  android apps: From research to practice,'' in \emph{Proceedings of the 39th
  International Conference on Software Engineering Companion}, ser. ICSE-C
  ’17.\hskip 1em plus 0.5em minus 0.4em\relax IEEE Press, 2017, p. 505–506.
  [Online]. Available: \url{https://doi.org/10.1109/ICSE-C.2017.166}
\BIBentrySTDinterwordspacing

\bibitem{Memon2013GUI}
A.~M. Memon and M.~B. Cohen, ``Automated testing of gui applications: Models,
  tools, and controlling flakiness,'' in \emph{Proceedings of the 2013
  International Conference on Software Engineering}, ser. ICSE ’13.\hskip 1em
  plus 0.5em minus 0.4em\relax IEEE Press, 2013, p. 1479–1480.

\bibitem{Coppola2017GUI}
\BIBentryALTinterwordspacing
R.~Coppola, M.~Morisio, and M.~Torchiano, ``Scripted gui testing of android
  apps: A study on diffusion, evolution and fragility,'' in \emph{Proceedings
  of the 13th International Conference on Predictive Models and Data Analytics
  in Software Engineering}, ser. PROMISE.\hskip 1em plus 0.5em minus
  0.4em\relax New York, NY, USA: Association for Computing Machinery, 2017, p.
  22–32. [Online]. Available: \url{https://doi.org/10.1145/3127005.3127008}
\BIBentrySTDinterwordspacing

\bibitem{CrashScope2017Moran}
\BIBentryALTinterwordspacing
K.~Moran, M.~Linares-V\'{a}squez, C.~Bernal-C\'{a}rdenas, C.~Vendome, and
  D.~Poshyvanyk, ``Crashscope: A practical tool for automated testing of
  android applications,'' in \emph{Proceedings of the 39th International
  Conference on Software Engineering Companion}, ser. ICSE-C ’17.\hskip 1em
  plus 0.5em minus 0.4em\relax IEEE Press, 2017, p. 15–18. [Online].
  Available: \url{https://doi.org/10.1109/ICSE-C.2017.16}
\BIBentrySTDinterwordspacing

\bibitem{Augusto2018M}
\BIBentryALTinterwordspacing
L.~Mariani, M.~Pezz\`{e}, and D.~Zuddas, ``Augusto: Exploiting popular
  functionalities for the generation of semantic gui tests with oracles,'' in
  \emph{Proceedings of the 40th International Conference on Software
  Engineering}, ser. ICSE ’18.\hskip 1em plus 0.5em minus 0.4em\relax New
  York, NY, USA: Association for Computing Machinery, 2018, p. 280–290.
  [Online]. Available: \url{https://doi.org/10.1145/3180155.3180162}
\BIBentrySTDinterwordspacing

\bibitem{GUI2019Denaro}
\BIBentryALTinterwordspacing
G.~Denaro, L.~Guglielmo, L.~Mariani, and O.~Riganelli, ``Gui testing in
  production: Challenges and opportunities,'' in \emph{Proceedings of the
  Conference Companion of the 3rd International Conference on Art, Science, and
  Engineering of Programming}, ser. Programming ’19.\hskip 1em plus 0.5em
  minus 0.4em\relax New York, NY, USA: Association for Computing Machinery,
  2019. [Online]. Available: \url{https://doi.org/10.1145/3328433.3328452}
\BIBentrySTDinterwordspacing

\bibitem{Practical2019Gu}
\BIBentryALTinterwordspacing
T.~Gu, C.~Sun, X.~Ma, C.~Cao, C.~Xu, Y.~Yao, Q.~Zhang, J.~Lu, and Z.~Su,
  ``Practical gui testing of android applications via model abstraction and
  refinement,'' in \emph{Proceedings of the 41st International Conference on
  Software Engineering}, ser. ICSE ’19.\hskip 1em plus 0.5em minus
  0.4em\relax IEEE Press, 2019, p. 269–280. [Online]. Available:
  \url{https://doi.org/10.1109/ICSE.2019.00042}
\BIBentrySTDinterwordspacing

\bibitem{Paladin}
\BIBentryALTinterwordspacing
Y.~Ma, Y.~Huang, Z.~Hu, X.~Xiao, and X.~Liu, ``Paladin: Automated generation of
  reproducible test cases for android apps,'' in \emph{Proceedings of the 20th
  International Workshop on Mobile Computing Systems and Applications}, ser.
  HotMobile ’19.\hskip 1em plus 0.5em minus 0.4em\relax New York, NY, USA:
  Association for Computing Machinery, 2019, p. 99–104. [Online]. Available:
  \url{https://doi.org/10.1145/3301293.3302363}
\BIBentrySTDinterwordspacing

\bibitem{Monkey}
A.~Developers, ``Ui/application exerciser monkey,'' 2012.

\bibitem{Wetzlmaier2017Hybrid}
\BIBentryALTinterwordspacing
T.~Wetzlmaier and R.~Ramler, ``Hybrid monkey testing: Enhancing automated gui
  tests with random test generation,'' in \emph{Proceedings of the 8th ACM
  SIGSOFT International Workshop on Automated Software Testing}, ser. A-TEST
  2017.\hskip 1em plus 0.5em minus 0.4em\relax New York, NY, USA: Association
  for Computing Machinery, 2017, p. 5–10. [Online]. Available:
  \url{https://doi.org/10.1145/3121245.3121247}
\BIBentrySTDinterwordspacing

\bibitem{Dynodroid}
\BIBentryALTinterwordspacing
A.~Machiry, R.~Tahiliani, and M.~Naik, ``Dynodroid: An input generation system
  for android apps,'' in \emph{Proceedings of the 2013 9th Joint Meeting on
  Foundations of Software Engineering}, ser. ESEC/FSE 2013.\hskip 1em plus
  0.5em minus 0.4em\relax New York, NY, USA: Association for Computing
  Machinery, 2013, p. 224–234. [Online]. Available:
  \url{https://doi.org/10.1145/2491411.2491450}
\BIBentrySTDinterwordspacing

\bibitem{Rico}
B.~Deka, Z.~Huang, C.~Franzen, J.~Hibschman, D.~Afergan, Y.~Li, J.~Nichols, and
  R.~Kumar, ``Rico: A mobile app dataset for building data-driven design
  applications,'' in \emph{Proceedings of the 30th Annual Symposium on User
  Interface Software and Technology}, ser. UIST '17, 2017.

\bibitem{liu2020owl}
\BIBentryALTinterwordspacing
Z.~Liu, C.~Chen, J.~Wang, Y.~Huang, J.~Hu, and Q.~Wang, ``Owl eyes: Spotting
  {UI} display issues via visual understanding,'' in \emph{35th {IEEE/ACM}
  International Conference on Automated Software Engineering, {ASE} 2020,
  Melbourne, Australia, September 21-25, 2020}.\hskip 1em plus 0.5em minus
  0.4em\relax {IEEE}, 2020, pp. 398--409. [Online]. Available:
  \url{https://doi.org/10.1145/3324884.3416547}
\BIBentrySTDinterwordspacing

\bibitem{wang2019iSENSE}
J.~Wang, Y.~Yang, R.~Krishna, T.~Menzies, and Q.~Wang, ``isense:
  Completion-aware crowdtesting management,'' in \emph{ICSE'2019}, 2019, pp.
  932--943.

\bibitem{wang2020context}
J.~Wang, Y.~Yang, S.~Wang, Y.~Hu, D.~Wang, and Q.~Wang, ``Context-aware
  in-process crowdworker recommendation,'' ser. ICSE 2020, 2020.

\bibitem{wang2021context}
J.~Wang, Y.~Yang, S.~Wang, C.~Chen, D.~Wang, and Q.~Wang, ``Context-aware
  personalized crowdtesting task recommendation,'' \emph{IEEE Transactions on
  Software Engineering}, 2021.

\bibitem{DBLP:journals/tse/Seaman99}
C.~B. Seaman, ``Qualitative methods in empirical studies of software
  engineering,'' \emph{{IEEE} Trans. Software Eng.}, vol.~25, no.~4, pp.
  557--572, 1999.

\bibitem{ren2015faster}
\BIBentryALTinterwordspacing
S.~Ren, K.~He, R.~B. Girshick, and J.~Sun, ``Faster {R-CNN:} towards real-time
  object detection with region proposal networks,'' \emph{{IEEE} Trans. Pattern
  Anal. Mach. Intell.}, vol.~39, no.~6, pp. 1137--1149, 2017. [Online].
  Available: \url{https://doi.org/10.1109/TPAMI.2016.2577031}
\BIBentrySTDinterwordspacing

\bibitem{he2016deep}
\BIBentryALTinterwordspacing
K.~He, X.~Zhang, S.~Ren, and J.~Sun, ``Deep residual learning for image
  recognition,'' in \emph{2016 {IEEE} Conference on Computer Vision and Pattern
  Recognition, {CVPR} 2016, Las Vegas, NV, USA, June 27-30, 2016}.\hskip 1em
  plus 0.5em minus 0.4em\relax {IEEE} Computer Society, 2016, pp. 770--778.
  [Online]. Available: \url{https://doi.org/10.1109/CVPR.2016.90}
\BIBentrySTDinterwordspacing

\bibitem{ResNet}
\BIBentryALTinterwordspacing
------, ``Deep residual learning for image recognition,'' in \emph{2016 IEEE
  Conference on Computer Vision and Pattern Recognition (CVPR)}.\hskip 1em plus
  0.5em minus 0.4em\relax Los Alamitos, CA, USA: IEEE Computer Society, jun
  2016, pp. 770--778. [Online]. Available:
  \url{https://doi.ieeecomputersociety.org/10.1109/CVPR.2016.90}
\BIBentrySTDinterwordspacing

\bibitem{ORB}
\BIBentryALTinterwordspacing
E.~Rublee, V.~Rabaud, K.~Konolige, and G.~Bradski, ``Orb: An efficient
  alternative to sift or surf,'' in \emph{Proceedings of the 2011 International
  Conference on Computer Vision}, ser. ICCV ’11.\hskip 1em plus 0.5em minus
  0.4em\relax USA: IEEE Computer Society, 2011, p. 2564–2571. [Online].
  Available: \url{https://doi.org/10.1109/ICCV.2011.6126544}
\BIBentrySTDinterwordspacing

\bibitem{SIFT}
\BIBentryALTinterwordspacing
D.~G. Lowe, ``Distinctive image features from scale-invariant keypoints,''
  \emph{Int. J. Comput. Vision}, vol.~60, no.~2, p. 91–110, Nov. 2004.
  [Online]. Available: \url{https://doi.org/10.1023/B:VISI.0000029664.99615.94}
\BIBentrySTDinterwordspacing

\bibitem{SURF}
\BIBentryALTinterwordspacing
H.~Bay, T.~Tuytelaars, and L.~Van~Gool, ``Surf: Speeded up robust features,''
  in \emph{Proceedings of the 9th European Conference on Computer Vision -
  Volume Part I}, ser. ECCV’06.\hskip 1em plus 0.5em minus 0.4em\relax
  Berlin, Heidelberg: Springer-Verlag, 2006, p. 404–417. [Online]. Available:
  \url{https://doi.org/10.1007/11744023_32}
\BIBentrySTDinterwordspacing

\bibitem{kotsiantis2007supervised}
S.~B. Kotsiantis, I.~Zaharakis, and P.~Pintelas, ``Supervised machine learning:
  A review of classification techniques,'' \emph{Emerging artificial
  intelligence applications in computer engineering}, vol. 160, pp. 3--24,
  2007.

\bibitem{berson2004overview}
A.~Berson, S.~Smith, and K.~Thearling, ``An overview of data mining
  techniques,'' \emph{Building Data Mining Application for CRM}, 2004.

\bibitem{breiman2001random}
L.~Breiman, ``Random forests,'' \emph{Machine learning}, vol.~45, no.~1, pp.
  5--32, 2001.

\bibitem{Deeplearning}
I.~Goodfellow, Y.~Bengio, and A.~Courville, \emph{Deep learning}.\hskip 1em
  plus 0.5em minus 0.4em\relax MIT press, 2016.

\bibitem{Deeplearningnature}
Y.~LeCun, Y.~Bengio, and G.~Hinton, ``Deep learning,'' \emph{nature}, vol. 521,
  no. 7553, pp. 436--444, 2015.

\bibitem{Easy-to-Deploy}
S.~{Ma}, Z.~{Xing}, C.~{Chen}, C.~{Chen}, L.~{Qu}, and G.~{Li},
  ``Easy-to-deploy api extraction by multi-level feature embedding and transfer
  learning,'' \emph{IEEE Transactions on Software Engineering}, pp. 1--1, 2019.

\bibitem{manning2008introduction}
C.~D. Manning, P.~Raghavan, and H.~Sch{\"u}tze, \emph{Introduction to
  information retrieval}.\hskip 1em plus 0.5em minus 0.4em\relax Cambridge
  university press, 2008.

\bibitem{DroidBot}
\BIBentryALTinterwordspacing
Y.~Li, Z.~Yang, Y.~Guo, and X.~Chen, ``Droidbot: A lightweight ui-guided test
  input generator for android,'' in \emph{Proceedings of the 39th International
  Conference on Software Engineering Companion}, ser. ICSE-C ’17.\hskip 1em
  plus 0.5em minus 0.4em\relax IEEE Press, 2017, p. 23–26. [Online].
  Available: \url{https://doi.org/10.1109/ICSE-C.2017.8}
\BIBentrySTDinterwordspacing

\bibitem{zein2016systematic}
S.~Zein, N.~Salleh, and J.~Grundy, ``A systematic mapping study of mobile
  application testing techniques,'' \emph{Journal of Systems and Software},
  vol. 117, pp. 334--356, 2016.

\bibitem{lamsa2017comparison}
T.~L{\"a}ms{\"a}, ``Comparison of gui testing tools for android applications,''
  2017.

\bibitem{reiss2018seeking}
S.~P. Reiss, Y.~Miao, and Q.~Xin, ``Seeking the user interface,''
  \emph{Automated Software Engineering}, vol.~25, no.~1, pp. 157--193, 2018.

\bibitem{behrang2018guifetch}
F.~Behrang, S.~P. Reiss, and A.~Orso, ``Guifetch: supporting app design and
  development through gui search,'' in \emph{Proceedings of the 5th
  International Conference on Mobile Software Engineering and Systems}.\hskip
  1em plus 0.5em minus 0.4em\relax ACM, 2018, pp. 236--246.

\bibitem{chen2020wireframe}
J.~Chen, C.~Chen, Z.~Xing, X.~Xia, L.~Zhu, J.~Grundy, and J.~Wang,
  ``Wireframe-based ui design search through image autoencoder,'' \emph{ACM
  Transactions on Software Engineering and Methodology (TOSEM)}, vol.~29,
  no.~3, pp. 1--31, 2020.

\bibitem{chen2019gallery}
C.~Chen, S.~Feng, Z.~Xing, L.~Liu, S.~Zhao, and J.~Wang, ``Gallery dc: Design
  search and knowledge discovery through auto-created gui component gallery,''
  \emph{Proceedings of the ACM on Human-Computer Interaction}, vol.~3, no.
  CSCW, pp. 1--22, 2019.

\bibitem{zhao2019actionnet}
D.~Zhao, Z.~Xing, C.~Chen, X.~Xia, and G.~Li, ``Actionnet: vision-based
  workflow action recognition from programming screencasts,'' in \emph{2019
  IEEE/ACM 41st International Conference on Software Engineering (ICSE)}.\hskip
  1em plus 0.5em minus 0.4em\relax IEEE, 2019, pp. 350--361.

\bibitem{yang2021uis}
B.~Yang, Z.~Xing, X.~Xia, C.~Chen, D.~Ye, and S.~Li, ``Uis-hunter: Detecting ui
  design smells in android apps,'' in \emph{2021 IEEE/ACM 43rd International
  Conference on Software Engineering: Companion Proceedings
  (ICSE-Companion)}.\hskip 1em plus 0.5em minus 0.4em\relax IEEE, 2021, pp.
  89--92.

\bibitem{yang2021don}
------, ``Don’t do that! hunting down visual design smells in complex uis
  against design guidelines,'' in \emph{2021 IEEE/ACM 43rd International
  Conference on Software Engineering (ICSE)}.\hskip 1em plus 0.5em minus
  0.4em\relax IEEE, 2021, pp. 761--772.

\bibitem{chen2021should}
Q.~Chen, C.~Chen, S.~Hassan, Z.~Xing, X.~Xia, and A.~E. Hassan, ``How should i
  improve the ui of my app? a study of user reviews of popular apps in the
  google play,'' \emph{ACM Transactions on Software Engineering and Methodology
  (TOSEM)}, vol.~30, no.~3, pp. 1--38, 2021.

\bibitem{nguyen2015reverse}
T.~A. Nguyen and C.~Csallner, ``Reverse engineering mobile application user
  interfaces with remaui (t),'' in \emph{Automated Software Engineering (ASE),
  2015 30th IEEE/ACM International Conference on}.\hskip 1em plus 0.5em minus
  0.4em\relax IEEE, 2015, pp. 248--259.

\bibitem{chen2018ui}
C.~Chen, T.~Su, G.~Meng, Z.~Xing, and Y.~Liu, ``From ui design image to gui
  skeleton: a neural machine translator to bootstrap mobile gui
  implementation,'' in \emph{Proceedings of the 40th International Conference
  on Software Engineering}.\hskip 1em plus 0.5em minus 0.4em\relax ACM, 2018,
  pp. 665--676.

\bibitem{moran2018machine}
\BIBentryALTinterwordspacing
K.~Moran, C.~Bernal{-}C{\'{a}}rdenas, M.~Curcio, R.~Bonett, and D.~Poshyvanyk,
  ``Machine learning-based prototyping of graphical user interfaces for mobile
  apps,'' \emph{{IEEE} Trans. Software Eng.}, vol.~46, no.~2, pp. 196--221,
  2020. [Online]. Available: \url{https://doi.org/10.1109/TSE.2018.2844788}
\BIBentrySTDinterwordspacing

\bibitem{chen2019storydroid}
S.~Chen, L.~Fan, C.~Chen, T.~Su, W.~Li, Y.~Liu, and L.~Xu, ``Storydroid:
  Automated generation of storyboard for android apps,'' in \emph{2019 IEEE/ACM
  41st International Conference on Software Engineering (ICSE)}.\hskip 1em plus
  0.5em minus 0.4em\relax IEEE, 2019, pp. 596--607.

\bibitem{chen2019gui}
S.~Chen, L.~Fan, C.~Chen, M.~Xue, Y.~Liu, and L.~Xu, ``Gui-squatting attack:
  Automated generation of android phishing apps,'' \emph{IEEE Transactions on
  Dependable and Secure Computing}, 2019.

\bibitem{feng2021auto}
S.~Feng, S.~Ma, J.~Yu, C.~Chen, T.~Zhou, and Y.~Zhen, ``Auto-icon: An automated
  code generation tool for icon designs assisting in ui development,'' in
  \emph{26th International Conference on Intelligent User Interfaces}, 2021,
  pp. 59--69.

\bibitem{zhao2021guigan}
T.~Zhao, C.~Chen, Y.~Liu, and X.~Zhu, ``Guigan: Learning to generate gui
  designs using generative adversarial networks,'' in \emph{2021 IEEE/ACM 43rd
  International Conference on Software Engineering (ICSE)}.\hskip 1em plus
  0.5em minus 0.4em\relax IEEE, 2021, pp. 748--760.

\bibitem{zhao2020seenomaly}
D.~Zhao, Z.~Xing, C.~Chen, X.~Xu, L.~Zhu, G.~Li, and J.~Wang, ``Seenomaly:
  vision-based linting of gui animation effects against design-don't
  guidelines,'' in \emph{2020 IEEE/ACM 42nd International Conference on
  Software Engineering (ICSE)}.\hskip 1em plus 0.5em minus 0.4em\relax IEEE,
  2020, pp. 1286--1297.

\bibitem{Chen2020From}
C.~Chunyang, F.~Sidong, L.~Zhengyang, X.~Zhenchang, Z.~Shengdong, and L.~Linda,
  ``From lost to found: Discover missing ui design semantics through recovering
  missing tags,'' in \emph{Proceedings of the ACM on Human-Computer
  Interaction, Volume. 4, No. CSCW, November 2020}, 2020.

\bibitem{Chen2020Object}
\BIBentryALTinterwordspacing
J.~Chen, M.~Xie, Z.~Xing, C.~Chen, X.~Xu, L.~Zhu, and G.~Li, ``Object detection
  for graphical user interface: old fashioned or deep learning or a
  combination?'' in \emph{{ESEC/FSE} '20: 28th {ACM} Joint European Software
  Engineering Conference and Symposium on the Foundations of Software
  Engineering, Virtual Event, USA, November 8-13, 2020}.\hskip 1em plus 0.5em
  minus 0.4em\relax {ACM}, 2020, pp. 1202--1214. [Online]. Available:
  \url{https://doi.org/10.1145/3368089.3409691}
\BIBentrySTDinterwordspacing

\bibitem{moran2018automated}
K.~Moran, B.~Li, C.~Bernal-C{\'a}rdenas, D.~Jelf, and D.~Poshyvanyk,
  ``Automated reporting of gui design violations for mobile apps,'' in
  \emph{Proceedings of the 40th International Conference on Software
  Engineering}.\hskip 1em plus 0.5em minus 0.4em\relax ACM, 2018, pp. 165--175.

\bibitem{moran2018detecting}
\BIBentryALTinterwordspacing
K.~Moran, C.~Watson, J.~Hoskins, G.~Purnell, and D.~Poshyvanyk, ``Detecting and
  summarizing {GUI} changes in evolving mobile apps,'' in \emph{Proceedings of
  the 33rd {ACM/IEEE} International Conference on Automated Software
  Engineering, {ASE} 2018, Montpellier, France, September 3-7, 2018},
  M.~Huchard, C.~K{\"{a}}stner, and G.~Fraser, Eds.\hskip 1em plus 0.5em minus
  0.4em\relax {ACM}, 2018, pp. 543--553. [Online]. Available:
  \url{https://doi.org/10.1145/3238147.3238203}
\BIBentrySTDinterwordspacing

\bibitem{chen2020unblind}
\BIBentryALTinterwordspacing
J.~Chen, C.~Chen, Z.~Xing, X.~Xu, L.~Zhu, G.~Li, and J.~Wang, ``Unblind your
  apps: predicting natural-language labels for mobile {GUI} components by deep
  learning,'' in \emph{{ICSE} '20: 42nd International Conference on Software
  Engineering, Seoul, South Korea, 27 June - 19 July, 2020}, G.~Rothermel and
  D.~Bae, Eds.\hskip 1em plus 0.5em minus 0.4em\relax {ACM}, 2020, pp.
  322--334. [Online]. Available: \url{https://doi.org/10.1145/3377811.3380327}
\BIBentrySTDinterwordspacing

\bibitem{lint}
\url{http://tools.android.com/tips/lint}, 2020.

\bibitem{chen2021accessible}
S.~Chen, C.~Chen, L.~Fan, M.~Fan, X.~Zhan, and Y.~Liu, ``Accessible or not an
  empirical investigation of android app accessibility,'' \emph{IEEE
  Transactions on Software Engineering}, 2021.

\bibitem{stylelint}
\url{https://github.com/stylelint/stylelint}, 2020.

\bibitem{gao2017every}
Y.~Gao, Y.~Luo, D.~Chen, H.~Huang, W.~Dong, M.~Xia, X.~Liu, and J.~Bu, ``Every
  pixel counts: Fine-grained ui rendering analysis for mobile applications,''
  in \emph{IEEE INFOCOM 2017-IEEE Conference on Computer Communications}.\hskip
  1em plus 0.5em minus 0.4em\relax IEEE, 2017, pp. 1--9.

\bibitem{li2019characterizing}
\BIBentryALTinterwordspacing
W.~Li, Y.~Jiang, C.~Xu, Y.~Liu, X.~Ma, and J.~Lu, ``Characterizing and
  detecting inefficient image displaying issues in android apps,'' in
  \emph{26th {IEEE} International Conference on Software Analysis, Evolution
  and Reengineering, {SANER} 2019, Hangzhou, China, February 24-27,
  2019}.\hskip 1em plus 0.5em minus 0.4em\relax {IEEE}, 2019, pp. 355--365.
  [Online]. Available: \url{https://doi.org/10.1109/SANER.2019.8668030}
\BIBentrySTDinterwordspacing

\bibitem{nayebi2012state}
F.~Nayebi, J.-M. Desharnais, and A.~Abran, ``The state of the art of mobile
  application usability evaluation,'' in \emph{2012 25th IEEE Canadian
  Conference on Electrical and Computer Engineering (CCECE)}.\hskip 1em plus
  0.5em minus 0.4em\relax IEEE, 2012, pp. 1--4.

\bibitem{holzinger2012making}
A.~Holzinger, P.~Treitler, and W.~Slany, ``Making apps useable on multiple
  different mobile platforms: On interoperability for business application
  development on smartphones,'' in \emph{International Conference on
  Availability, Reliability, and Security}.\hskip 1em plus 0.5em minus
  0.4em\relax Springer, 2012, pp. 176--189.

\bibitem{white2019improving}
T.~D. White, G.~Fraser, and G.~J. Brown, ``Improving random gui testing with
  image-based widget detection,'' in \emph{Proceedings of the 28th ACM SIGSOFT
  International Symposium on Software Testing and Analysis}.\hskip 1em plus
  0.5em minus 0.4em\relax ACM, 2019, pp. 307--317.

\bibitem{degott2019learning}
C.~Degott, N.~P. Borges~Jr, and A.~Zeller, ``Learning user interface element
  interactions,'' in \emph{Proceedings of the 28th ACM SIGSOFT International
  Symposium on Software Testing and Analysis}.\hskip 1em plus 0.5em minus
  0.4em\relax ACM, 2019, pp. 296--306.

\bibitem{zhang2015compatibility}
T.~Zhang, J.~Gao, J.~Cheng, and T.~Uehara, ``Compatibility testing service for
  mobile applications,'' in \emph{2015 IEEE Symposium on Service-Oriented
  System Engineering}.\hskip 1em plus 0.5em minus 0.4em\relax IEEE, 2015, pp.
  179--186.

\bibitem{ki2019mimic}
T.~Ki, C.~M. Park, K.~Dantu, S.~Y. Ko, and L.~Ziarek, ``Mimic: Ui compatibility
  testing system for android apps,'' in \emph{2019 IEEE/ACM 41st International
  Conference on Software Engineering (ICSE)}.\hskip 1em plus 0.5em minus
  0.4em\relax IEEE, 2019, pp. 246--256.

\bibitem{kondratova2011culturally}
I.~Kondratova and I.~Goldfarb, ``Culturally appropriate web user interface
  design study: Research methodology and results,'' in \emph{Handbook of
  research on culturally-aware information technology: Perspectives and
  models}.\hskip 1em plus 0.5em minus 0.4em\relax IGI Global, 2011, pp.
  316--336.

\bibitem{mahajan2015detection}
S.~Mahajan and W.~G. Halfond, ``Detection and localization of html presentation
  failures using computer vision-based techniques,'' in \emph{2015 IEEE 8th
  International Conference on Software Testing, Verification and Validation
  (ICST)}.\hskip 1em plus 0.5em minus 0.4em\relax IEEE, 2015, pp. 1--10.

\bibitem{mahajan2015websee}
------, ``Websee: A tool for debugging html presentation failures,'' in
  \emph{2015 IEEE 8th International Conference on Software Testing,
  Verification and Validation (ICST)}.\hskip 1em plus 0.5em minus 0.4em\relax
  IEEE, 2015, pp. 1--8.

\bibitem{mahajan2016using}
S.~Mahajan, B.~Li, P.~Behnamghader, and W.~G. Halfond, ``Using visual symptoms
  for debugging presentation failures in web applications,'' in \emph{2016 IEEE
  International Conference on Software Testing, Verification and Validation
  (ICST)}.\hskip 1em plus 0.5em minus 0.4em\relax IEEE, 2016, pp. 191--201.

\bibitem{saar2016browserbite}
T.~Saar, M.~Dumas, M.~Kaljuve, and N.~Semenenko, ``Browserbite: cross-browser
  testing via image processing,'' \emph{Software: Practice and Experience},
  vol.~46, no.~11, pp. 1459--1477, 2016.

\bibitem{DBLP:conf/icse/MahajanAMH18}
\BIBentryALTinterwordspacing
S.~Mahajan, N.~Abolhassani, P.~McMinn, and W.~G.~J. Halfond, ``Automated repair
  of mobile friendly problems in web pages,'' in \emph{Proceedings of the 40th
  International Conference on Software Engineering, {ICSE} 2018, Gothenburg,
  Sweden, May 27 - June 03, 2018}.\hskip 1em plus 0.5em minus 0.4em\relax
  {ACM}, 2018, pp. 140--150. [Online]. Available:
  \url{https://doi.org/10.1145/3180155.3180262}
\BIBentrySTDinterwordspacing

\bibitem{DBLP:conf/apsec/MahajanGPH16}
\BIBentryALTinterwordspacing
S.~Mahajan, K.~B. Gadde, A.~Pasala, and W.~G.~J. Halfond, ``Detecting and
  localizing visual inconsistencies in web applications,'' in \emph{23rd
  Asia-Pacific Software Engineering Conference, {APSEC} 2016, Hamilton, New
  Zealand, December 6-9, 2016}.\hskip 1em plus 0.5em minus 0.4em\relax {IEEE}
  Computer Society, 2016, pp. 361--364. [Online]. Available:
  \url{https://doi.org/10.1109/APSEC.2016.060}
\BIBentrySTDinterwordspacing

\bibitem{DBLP:conf/issta/MahajanAMH17a}
\BIBentryALTinterwordspacing
S.~Mahajan, A.~Alameer, P.~McMinn, and W.~G.~J. Halfond, ``Xfix: an automated
  tool for the repair of layout cross browser issues,'' in \emph{Proceedings of
  the 26th {ACM} {SIGSOFT} International Symposium on Software Testing and
  Analysis, Santa Barbara, CA, USA, July 10 - 14, 2017}, T.~Bultan and K.~Sen,
  Eds.\hskip 1em plus 0.5em minus 0.4em\relax {ACM}, 2017, pp. 368--371.
  [Online]. Available: \url{https://doi.org/10.1145/3092703.3098223}
\BIBentrySTDinterwordspacing

\bibitem{mahajan2021effective}
S.~Mahajan, A.~Alameer, P.~McMinn, and W.~G. Halfond, ``Effective automated
  repair of internationalization presentation failures in web applications
  using style similarity clustering and search-based techniques,''
  \emph{Software Testing, Verification and Reliability}, vol.~31, no. 1-2, p.
  e1746, 2021.

\bibitem{alameer2019efficiently}
A.~Alameer, P.~T. Chiou, and W.~G. Halfond, ``Efficiently repairing
  internationalization presentation failures by solving layout constraints,''
  in \emph{2019 12th IEEE Conference on Software Testing, Validation and
  Verification (ICST)}.\hskip 1em plus 0.5em minus 0.4em\relax IEEE, 2019, pp.
  172--182.

\bibitem{DBLP:conf/icst/AlthomaliKM19}
\BIBentryALTinterwordspacing
I.~Althomali, G.~M. Kapfhammer, and P.~McMinn, ``Automatic visual verification
  of layout failures in responsively designed web pages,'' in \emph{12th {IEEE}
  Conference on Software Testing, Validation and Verification, {ICST} 2019,
  Xi'an, China, April 22-27, 2019}.\hskip 1em plus 0.5em minus 0.4em\relax
  {IEEE}, 2019, pp. 183--193. [Online]. Available:
  \url{https://doi.org/10.1109/ICST.2019.00027}
\BIBentrySTDinterwordspacing

\end{thebibliography}

\newpage

\end{document}